\documentclass[journal,comsoc]{IEEEtran}

\usepackage{amsfonts}
\IEEEoverridecommandlockouts
\ifCLASSINFOpdf
\else
\fi
\usepackage{booktabs}
\usepackage{makecell}
\usepackage{graphicx}

\usepackage{subfigure}
\usepackage{cite}
\usepackage{color}
\usepackage[pagebackref=false,breaklinks=false,letterpaper=true,colorlinks,citecolor=blue,linkcolor=blue, anchorcolor=blue, bookmarks=false]{hyperref}
\usepackage{amsmath}
\usepackage{amssymb}
\usepackage{array}
\usepackage{bm}
\usepackage{algorithm}
\usepackage{algorithmic}
\usepackage{pifont}
\newcommand{\et}{\emph{et al.}}

\usepackage{cite}
\usepackage{lettrine} 

\title{SSwsrNet: A Semi-Supervised Few-Shot Learning Framework for Wireless Signal Recognition}
\author{Hao Zhang, \IEEEmembership{Graduate Student Member, IEEE}, Fuhui Zhou, \IEEEmembership{Senior Member, IEEE}, 
\\Qihui Wu, \IEEEmembership{Fellow, IEEE}, and Naofal Al-Dhahir, \IEEEmembership{Fellow, IEEE}
\thanks{
This work was supported by the National Natural Science Foundation of China under Grant 62031012 and Grant 62222107, the China Scholarship Council under Grant 202306830108, the Postgraduate Research \& Practice Innovation Program of Jiangsu Province under Grant KYCX23\_0380, and the Interdisciplinary Innovation Fund for Doctoral Students of Nanjing University of Aeronautics and Astronautics under Grant KXKCXJJ202302. 
The work of N. Al-Dhahir was supported by Erik Jonson Distinguished Professorship at UT-Dallas.
(\emph{Corresponding author: Fuhui Zhou.})

Hao Zhang, Fuhui Zhou, and Qihui Wu are with the College of Electronic and Information Engineering, Nanjing University of Aeronautics and Astronautics, Nanjing 211106 China. They are also with the Key Laboratory of Dynamic Cognitive System of Electromagnetic Spectrum Space (Nanjing University of Aeronautics and Astronautics) and with the Ministry of Industry and Information Technology, Nanjing, 211106, China (email: haozhangcn@nuaa.edu.cn, zhoufuhui@ieee.org, wuquhui2014@sina.com).

Naofal Al-Dhahir is with the Department of Electrical and Computer Engineering, The University of Texas at Dallas, Richardson, TX 75080 USA (e-mail: aldhahir@utdallas.edu).
}
}

\begin{document}

\maketitle

\begin{abstract}
Wireless signal recognition (WSR) is crucial in modern and future wireless communication networks since it aims to identify properties of the received signal. Although many deep learning-based WSR models have been developed, they still rely on a large amount of labeled training data. Thus, they cannot tackle the few-sample problem in the practically and dynamically changing wireless communication environment. To overcome this challenge, a novel SSwsrNet framework is proposed by using the deep residual shrinkage network (DRSN) and semi-supervised learning. The DRSN can learn discriminative features from noisy signals. Moreover, a modular semi-supervised learning method that combines labeled and unlabeled data using MixMatch is exploited to further improve the classification performance under few-sample conditions. Extensive simulation results on automatic modulation classification (AMC) and wireless technology classification (WTC) demonstrate that our proposed WSR scheme can achieve better performance than the benchmark schemes in terms of classification accuracy. This novel method enables more robust and adaptive signal recognition for next-generation wireless networks. 
\end{abstract}


\section{Introduction}
\IEEEPARstart{W}{ITH} the rapidly growing number of devices in wireless communication systems, spectrum resources are becoming increasingly scarce. To efficiently utilize spectrum resources, especially the underutilized frequency bands, dynamic spectrum access (DSA) technology has been introduced for cognitive radio systems \cite{wu2014cognitive,ding2015cellular}. However, DSA-enabled cognitive radio systems cannot dynamically avoid interference when sharing spectrum with unlicensed bands. Therefore, cognitive radios must ensure that they do not interfere with transmissions from incumbent users. To avoid interference with incumbent users, it is vital to distinguish their transmissions from those of secondary users. This can be achieved by signal recognition algorithms that classify incoming transmissions. In this case, signal recognition can be utilized to enhance the spectrum efficiency and security of spectrum sharing. 

Wireless signal recognition (WSR) involves characterizing the received signals in non-cooperative communication environments without prior knowledge of the transmission parameters. WSR has numerous applications across military and civilian domains, including decoding of intercepted transmissions, adaptive demodulation, and DSA. Wireless signal recognition can be classified into several primary applications, including automatic modulation classification (AMC), wireless technology classification (WTC)/wireless standard identification (WSI), and radio frequency (RF) fingerprinting identification. AMC aims to differentiate various modulation types such as amplitude modulation (AM) and frequency modulation (FM), phase shift keying (PSK), or quadrature amplitude modulation (QAM), while the objective of WTC is to recognize the wireless standard of an unidentified signal, such as 4G LTE or IEEE 802.11. WTC is more challenging than AMC since a wireless standard typically encompasses multiple modulation types that may vary dynamically based on factors, such as the channel quality and the transmit scheme. 

\subsection{Related Works and Motivations}
Wireless signal recognition (WSR) techniques can be categorized into model-driven (classical) and data-driven methods \cite{zhang2021novel}. Classical WSR predominantly leverages likelihood-based (LB) methods and feature-based (FB) methods. LB techniques, such as the average likelihood ratio test (ALRT) \cite{huan1995likelihood,beidas1995higher}, the generalized likelihood ratio test (GLRT) \cite{panagiotou2000likelihood}, and the hybrid likelihood ratio test (HLRT) \cite{ozdemir2015asynchronous} formulate WSR as a hypothesis testing problem by the likelihood functions, which require known signal-to-noise ratio (SNR) and frequency offsets. However, these parameters are often unavailable in non-cooperative conditions, limiting practical LB applicability. FB methods instead manually extract statistical features, such as moments \cite{moser2015automatic}, cumulants \cite{li2021modulation}, and cyclostationarity from received signals to train classifiers, such as logistic regression \cite{jiang2018feature}, support vector machines \cite{cai2012generalized}, and k-nearest neighbors \cite{yang2019adaptive}. However, their performance strongly depends on hand-crafted features, which are susceptible to noise and have intrinsic limitations. Overall, although the classical WSR techniques have been widely studied, their reliance on expert feature engineering and prior channel knowledge constrains the achievable accuracy and robustness. This has motivated growing interest in data-driven deep learning approaches that directly learn features from raw signal samples and channel effects.

Deep learning has made great progress in many areas, especially in computer vision \cite{wang2021small,xu2020automatic}, natural language processing, as well as wireless communications \cite{zhang2021novel,zhang2021automatic}. Deep learning holds promise for advancing WSR by learning discriminative features directly from raw signal data in an end-to-end manner. It can automatically extract features for a classification task through backpropagation, outperforming hand-crafted features used in the classical methods. However, the performance of WSR methods based on deep learning strongly depends on the availability of large labeled datasets, which can be prohibitively expensive or infeasible to collect in wireless communication domains. 
The non-cooperative nature of WSR makes it extremely challenging to obtain extensive labeled signal data. 
Firstly, wireless signals are often transmitted in a non-cooperative manner, which means that labeled data collection requires extensive time, resources, and expertise. 
Secondly, the wireless channel is dynamic and time-varying, which makes it difficult to collect enough labeled data in a static environment, where the collected data is effective. 
Thirdly, the wireless signal is sensitive to the interference, which makes it difficult to collect data with high-quality labels in a complex environment, where the label quality will significantly affect the performance of the model. 
Therefore, developing deep learning techniques that can learn from finite labeled examples supplemented by abundant unlabeled data through semi-supervised or transfer learning is an important research direction. Recent works have explored leveraging unlabeled samples for data augmentation, pre-training, and regularization. However, the reliable application of deep learning for WSR with restricted labeled data remains an open problem. 

Deep learning, especially convolutional neural networks (CNNs) and recurrent neural networks (RNNs) have been extensively studied for WSR. Early work by O'Shea \et \cite{o2016convolutional} adopted a simple two-layer CNN architecture followed by a softmax classifier. Later, more advanced CNN models were proposed by incorporating insights from computer vision, such as residual learning in ResNet \cite{he2016deep} to mitigate vanishing gradients when training deep networks. Direct application of ResNet for AMC yielded limited gains, motivating custom architectures such as the IQ-ResNet \cite{o2018over} which operates on complex-valued signals. Other works sought to optimize efficiency through models, such as the Involutional ResNet \cite{zhang2021automatic}, balancing accuracy and inference speed. Moreover, multi-scale convolutional networks have been applied for both AMC and WTC \cite{zhang2021novel,yuan2021multiscale}. Meanwhile, RNN architectures including long short-term memory (LSTM) \cite{hu2019deep} and gated recurrent units (GRU) \cite{huang2020automatic} have been explored to leverage timing correlations in signals. However, RNNs entail high training costs and are often constrained to fewer modulation types. Beyond basic CNNs and RNNs, researchers have further advanced AMC via techniques, such as feature fusion \cite{zhang2019automatic}, adversarial training \cite{zhang2020automatic}, and joint data and knowledge transfer \cite{ding2022data}. The latter work notably combined data-driven learning with expert wireless knowledge to extract both statistical and deep representations \cite{ding2022data}. 
Recently, novel MC techniques \cite{ryu2023emc} and efficient networks \cite{zhang2023amc} have been proposed for WSR to enhance the recognition performance. 
In summary, while CNNs and RNNs have driven initial wireless deep learning research, a key limitation that remains is their reliance on large labeled datasets for training. 

\begin{figure}
\centering
\subfigure[]{
\includegraphics[width=0.95\linewidth]{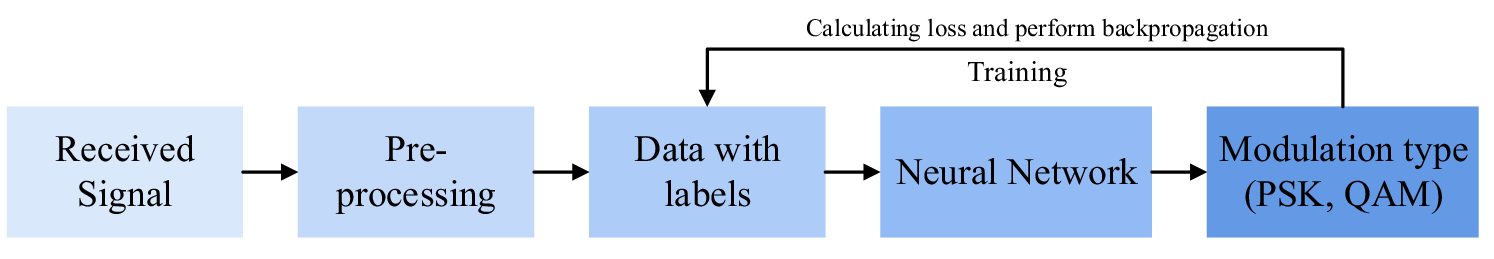} 
}
\subfigure[]{
\includegraphics[width=0.95\linewidth]{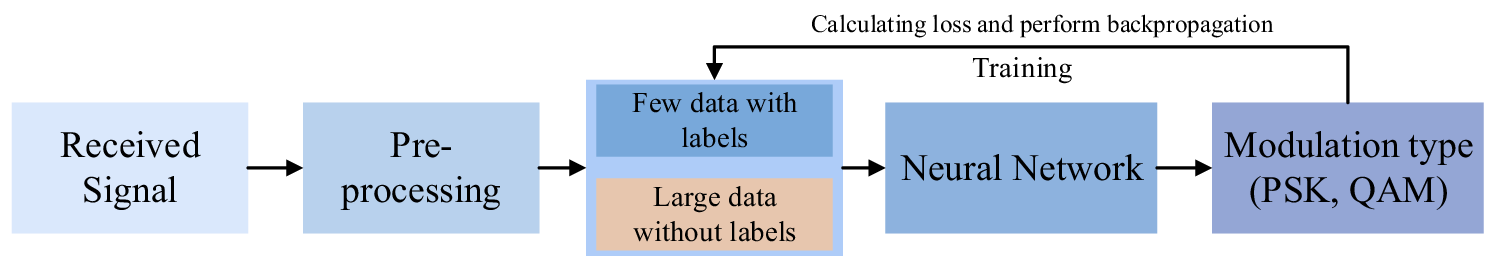} 
}
\subfigure[]{
\includegraphics[width=0.95\linewidth]{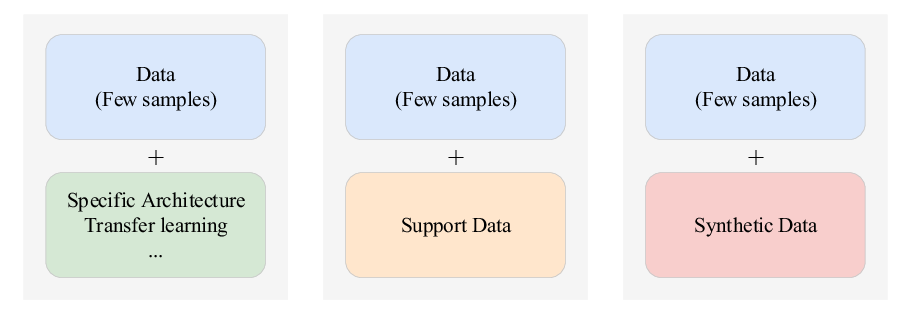} 
}
\DeclareGraphicsExtensions.
\caption{The structure of a DL-based WSR method. (a) common DL-based method; (b) few-sample conditions; (c) WSR methods under few-sample conditions.}
\label{fig:moti}
\end{figure}

For WSR under few-sample conditions, there have been several recent works. As shown in Fig. \ref{fig:moti}, these recent works can be classified into three categories, namely few-sample-based methods, support-data-based methods, and synthetic-data-based methods. For few-sample-based methods, they only utilize the labeled data with few samples by using specifically designed architectures and transfer learning. 
Li \et \cite{li2020automatic} introduced a neural architecture termed AMR-CapsNet for AMC that achieves high accuracy under limited training data. 
The authors in \cite{bu2020adversarial} proposed a transfer learning method for AMC. 
For support-data-based methods, the proposed methods are first pre-trained on a specific dataset named support set to learn the generative learning ability for WSR. Then, the pre-trained model is trained on the few-sample conditions, namely, the few-shot learning stage. The support set is often selected from the same dataset, where the classes cannot appear in the few-shot learning stage. For WSR, it is not applicable to collect enough additional labeled data to construct the support set under non-cooperative conditions. 
Zhou \et \cite{zhou2021amcrn} proposed a new network, AMC relation network (AMCRN), to boost performance under few-shot conditions. 
For synthetic-data-based methods, it is intuitive to “generate” more labeled data for the training using a generative adversarial network (GAN). 
To address the limited availability of labeled training data for AMC, Zhang \et \cite{zhang2022gan} proposed a new few-shot learning framework by using GAN and SNR augmentation to expand the diversity of the training set. 
These works have made impressive progress under the condition of fewer samples with the help of the few-shot learning. However, these few-shot learning methods are applicable for the AMC applications, especially under non-cooperative conditions, where labeled data are difficult to obtain. Fortunately, there are large amounts of unlabeled data that can be utilized without any constraints. Moreover, the modulation signals from the system without labels can be further utilized for improving the classification performance. 
Thus, in this research, we investigate using both labeled and unlabeled data simultaneously to solve the problems of using few samples in the WSR task.

\subsection{Contributions}

The nature of wireless communications, particularly the non-cooperative, dynamic and open characteristics, makes it extremely challenging to collect extensive labeled signal data. 
The few-shot learning can be utilized to address the challenge of the limited labeled data for WSR by using few-sample data. 
However, when training with few samples, the trained model has poor generalization ability and cannot be applied for other datasets. Moreover, models may easily overfit the few training data. 
How to effectively apply few-shot learning for WSR and improve the generalization ability under few-sample conditions is a critical issue for WSR. 
Intuitively, the unlabeled data can be utilized to improve the classification performance under few-sample conditions by using semi-supervised learning. 
However, when the labeled data is limited, noise in the data may significantly affect the performance of the model and the model may not be robust to the noise. 
To mitigate the reliance on large labeled datasets for deep learning-based WSR, this paper proposes a novel model termed SSwsrNet by combining a deep residual shrinkage network (DRSN) with the semi-supervised learning via MixMatch. DRSN is designed to robustly extract discriminative features from noisy signals. Meanwhile, MixMatch leverages both labeled and unlabeled data during training to improve the generalization under limited supervised samples. 
The key contributions are developing a tailored deep architecture for wireless feature extraction and advancing semi-supervised learning for the wireless communications domain. SSwsrNet provides an important step toward reducing reliance on large labeled datasets for applying deep neural networks to real-world wireless applications. The contributions of this paper can be summarized as follows. 
\begin{enumerate}
	\item The existing DL-based WSR models cannot simultaneously utilize labeled and unlabeled data under few-sample conditions. It is the first time that both labeled and unlabeled data are exploited in the WSR training process for tackling less samples under few-sample conditions. For few-sample recognition, we propose a novel SSwsrNet framework with a deep residual shrinkage network (DRSN) and a modular semi-supervised learning method named MixMatch. DRSN can learn more discriminative features from noisy signals than the state-of-the-art DL-based schemes. The semi-supervised learning MixMatch is motivated by the idea of data augmentation, and it can be flexibly integrated into other DL-based schemes.
	
	\item To learn discriminative features without irrelevant noise information, a deep residual shrinkage network with residual stacks is proposed for WSR. To enable the model with the ability to learn deep features without the gradient exploding problem, the residual stacks constructed by residual shrinkage units (RSU) with shrinkage block and identity shortcut are developed. The shrinkage block is designed to learn discriminative features from high-dimensional tensors by using several convolutional and FC layers. Moreover, soft thresholding can help the model mitigate the influence of noise from the raw input signals. 
	
	\item 
	Extensive simulation results demonstrate the superiority of our proposed WSR framework SSwsrNet compared with other DL-based WSR schemes in terms of the classification accuracy. Moreover, the visualized features present the discriminative and separable representation ability and the superiority of our proposed WSR framework from another aspect. 
\end{enumerate}

The remainder of this paper is organized as follows. Section \ref{sec:signal_model} provides background about the wireless signal model and fundamental DL techniques for WSR. Section \ref{sec:proposed_drsn} introduces the proposed SSwsrNet framework by combining deep residual shrinkage networks with modular semi-supervised learning. Section \ref{sec:experiments} presents experimental results. Finally, Section \ref{sec:conclusion} concludes the paper. The list of abbreviations is presented in Table \ref{tab:abbrv}.

\begin{table}[h]
	\centering
	\caption{List of Abbreviations.}
	\begin{center}
	\begin{tabular}{m{2cm}<{\centering}m{5cm}<{\centering}}
	\toprule
	Abbreviations & Description \\
	\midrule
	AM & amplitude modulation\\
	AMC & automatic modulation classification\\
	AWGN & additive white Gaussian noise\\
	DL & deep learning\\
	DSA & dynamic spectrum access\\
	DRSN & deep residual shrinkage network\\
	FB & feature-based\\
	FC & fully connected\\
	FM & frequency modulation\\
	GAN & generative adversarial network\\
	GAP & global average pooling\\
	IQ &  in-phase and quadrature\\
	LB & likelihood-based\\
	LSTM & long short-term memory\\
	PSK & phase shift keying\\
	QAM	& quadrature amplitude modulation\\
	RSU & residual shrinkage unit\\
	SNR & signal-to-noise ratio\\
	WSR & wireless signal recognition\\
	WTC & wireless technology classification \\
	\bottomrule
	\end{tabular}
	\label{tab:abbrv}
	\end{center}
\end{table}

\section{Signal Model}
\label{sec:signal_model}
According to the previous work \cite{zhang2021novel}, a single antenna is used to receive the signal. 
The received signal with the $c$th modulation hypothesis $H_c$ can be expressed as
\begin{equation}
H_c: x(l)=s_c(l)+n(l), l=1, 2, \cdots, L,
\end{equation}
where $c$ denotes the index of the modulation hypothesis. 
$l$ is the sample index, while $L$ is the maximum number of samples. 
Besides, the noise term $n(l)$ is typically modeled as additive white Gaussian noise (AWGN) with zero mean and variance $\sigma_\omega^2$. The signal-to-noise ratio (SNR) is then defined as $\gamma=|h|^2/\sigma_\omega^2$, where $|h|^2$ is the channel gain. 

At the receiver, which has a single antenna, the received data is processed by sampling and quantizing to obtain the complex-valued raw signal symbols. These raw signal symbols are then transformed into in-phase (I) and quadrature (Q) sequences, expressed as
\begin{equation}
\textbf{x}^{IQ}=\begin{pmatrix}
\Re[x(1),\cdots,x(L)] \\
\Im[x(1),\cdots,x(L)]
\end{pmatrix},
\end{equation}
where $\Re$ and $\Im$ denote the real and imaginary parts of the signal, respectively. 

After pre-processing, the raw signal symbols are transformed from $\textbf{x} \in \mathbb{R}^{1\times L}$ to $\textbf{x}^{IQ} \in \mathbb{R}^{2\times L}$. Then, the outputs are utilized as the input data of the SSwsrNet framework, and the feature extraction process can be expressed as 
\begin{equation}
	f_e: \textbf{x}^{IQ} \in \mathbb{R}^{2\times L} \to \textbf{x}_e \in \mathbb{R}^{1\times i},
\end{equation}
where $\textbf{x}_e$ denotes the outputs of the feature extractor, and $i$ is the size of the kernel. 

The learned features are then classified by the classification module, comprised of fully connected (FC) layers followed by a softmax classifier. The final layer consists of an FC layer with the number of neurons equal to the number of modulation types. Thus, each neuron corresponds to one modulation type, and the softmax activation function converts the outputs to probability values representing the likelihood that the input belongs to each candidate modulation format. The softmax function can be expressed as
\begin{equation}
	Softmax(z_i)=\frac{\exp(z_i)}{\sum_{c=1}^C \exp{z_c}},
\end{equation}
where $z_i$ is the output of the $i$th neuron, and $C$ is the total number of neurons, which is equal to the number of classes.

\section{SSwsrNet: Our Proposed WSR Framework}
\label{sec:proposed_drsn}

In this section, we describe the details of our proposed SSwsrNet framework for wireless signal recognition (WSR). 
First, the overall framework is shown in Fig. \ref{fig:model}. 
Next, the feature extraction module and the modular semi-supervised module are introduced, respectively. 
Finally, training strategies for the proposed framework are described. 

\subsection{Overview of the Proposed SSwsrNet Framework}

\begin{figure*}
\centering
\includegraphics[width=0.95\linewidth]{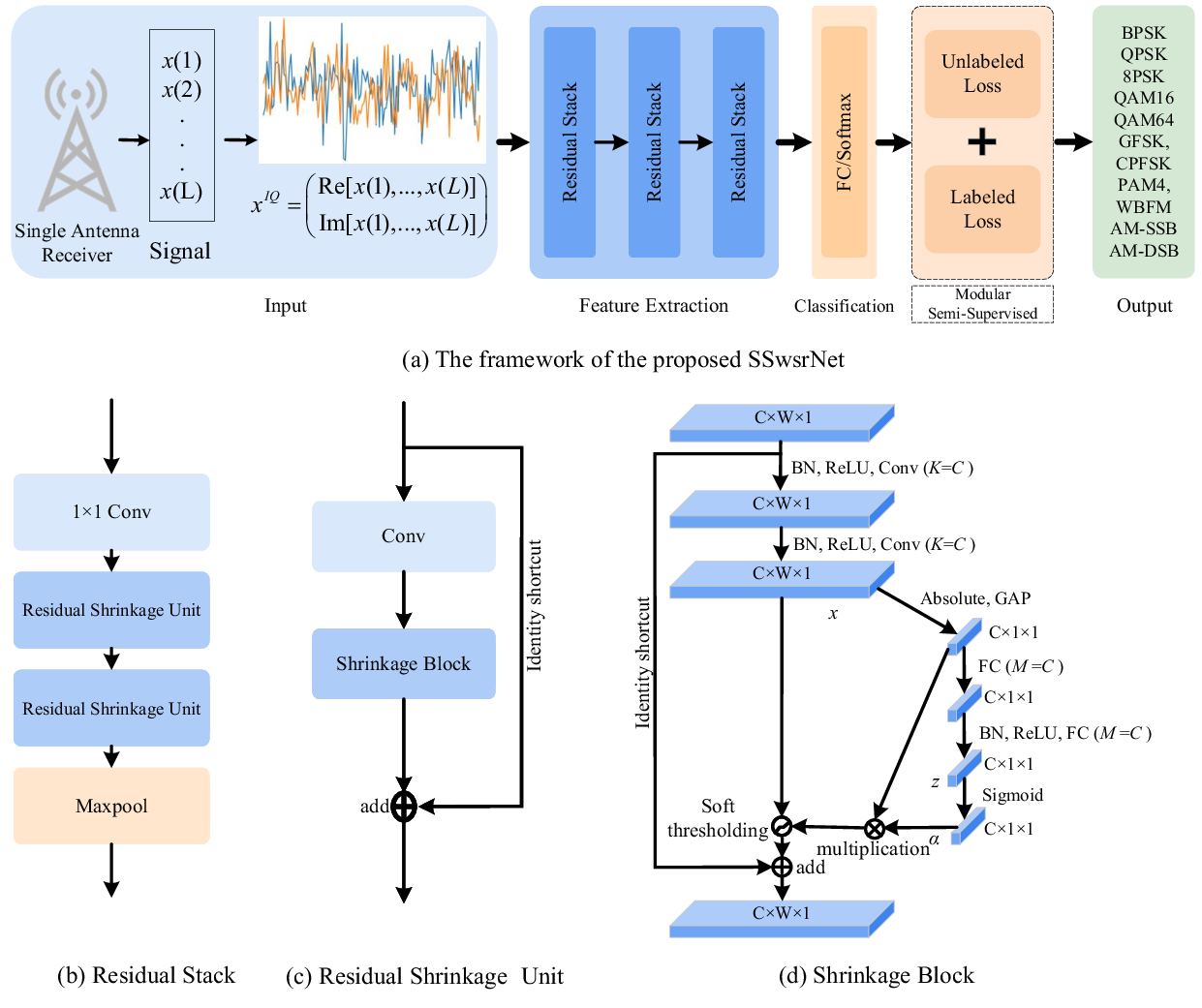} 
\caption{The structure of the proposed SSwsrNet. }
\label{fig:model}
\end{figure*}

In this section, we present the details of our proposed SSwsrNet framework. As illustrated in Fig. \ref{fig:model}, the framework of the proposed SSwsrNet comprises four main components, namely, the input layer, the feature extraction module, the classification module, and the modular semi-supervised learning module (only applicable when training). First, in the input layer, the received signal samples are transformed into in-phase and quadrature (IQ) samples to match the input requirements of the model. Next, a deep residual shrinkage network (DRSN) is utilized to extract discriminative features from the input signals. The DRSN incorporates the proposed residual shrinkage modules to capture salient features from the raw data. Subsequently, in the classification stage, the cross-entropy loss (\emph{i.e.} the softmax loss from the final fully connected (FC) layer) is employed to optimize the model parameters. Furthermore, a modular semi-supervised learning approach is utilized to enable classification under few-shot conditions during training. Specifically, the top layer of the SSwsrNet is meta-trained as low-density separators on an unlabeled dataset to improve the generalizability of the top layer units. The model is then fine-tuned on the small labeled dataset.

\subsection{Feature Extraction Module}
In this work, we propose a DRSN to learn highly discriminative features, aiming to improve the classification accuracy of WSR. As shown in Fig. \ref{fig:model}, the proposed DRSN consists of three residual stack modules, an FC layer and the softmax classifier. In DRSN, the residual stack module shares the same structure as its conventional version in \cite{o2018over}. However, it is more powerful for capturing accurate information mainly due to the design of the residual unit. 
Instead of using two convolutional layers in the original residual unit, we replace the second convolutional layer with the advanced shrinkage block \cite{zhao2019deep} to make it more robust to noise. 
By inheriting both the advantages of the residual units and shrinkage blocks, the proposed DRSN can learn more accurate features from the input signals. 
Firstly, the residual unit can effectively alleviate the gradient vanishing problem, enabling the network to effectively capture deeper and more complex features of the input data, improving accuracy and robustness of information extraction. 
Secondly, the shrinkage block can learn more discriminative features from the input signals by adaptively determining the threshold based on the data, making the process more flexible and data-driven.

\subsubsection{Residual Stack} 
To evaluate the efficacy of the proposed residual shrinkage module for WSR, the residual stack shares the same structure as the original ResNet in \cite{o2018over}. As shown in Fig. \ref{fig:model} (b), the residual stack starts with a convolutional layer with a $1\times 1$ kernel, which is mainly used to change the channel number and aggregate information from the previous layer. Then, two residual units are utilized together to extract the features. Finally, a max-pooling operation is utilized to decrease the width of the feature maps and to reduce the computational cost for subsequent layers. In contrast to using a single residual unit as in \cite{west2017deep}, the proposed residual stack structure with double residual units can achieve significant gains in terms of the classification accuracy. Thus, the elegant structure is adopted to inherit the advantages of residual units.

\subsubsection{Residual Shrinkage Unit}

As illustrated in Fig. \ref{fig:model} (c), the proposed residual shrinkage unit (RSU) is a modified residual unit incorporating a channel-wise soft thresholding operation \cite{zhao2019deep}. Specifically, the input feature map $x$ is first flattened into a 1D vector by using an absolute value operation followed by the global average pooling (GAP). This vector is then passed through a two-layer FC network. The second FC layer contains multiple neurons equal to the number of channels in $x$. The output of this FC network is scaled to the range $(0,1)$ by using a sigmoid activation, given as
\begin{equation}
	\alpha_c=\frac{1}{1+\exp(-z_c)},
\end{equation}
where $z_c$ denotes the output of the $c$th neuron in the FC layer, and $\alpha_c$ represents the $c$th learned scaling parameter. The threshold for each channel is then computed as 
\begin{equation}
	\tau_c=\alpha_c \underset{i,j}{\mathrm{average} } |x_{i,j,c}|,
\end{equation}
where $\tau_c$ represents the threshold for the $c$th channel and $i$, $j$, and $c$ denote width, height, and channel indexes of the input feature map $x$, respectively, and $\mathrm{average}$ denotes the average operation. By stacking multiple RSUs, the model can learn highly discriminative features through successive nonlinear transformations. The soft thresholding acts as a shrinkage function within each RSU to suppress noise-related information. The layered RSUs enable deep feature learning while enhancing the discriminability of the features. 

GAP computes the mean across each channel of the feature map \cite{lin2013network}. It is commonly placed before the final output layer in CNN architectures. By reducing the spatial dimensions to a single value per channel, GAP drastically decreases the number of parameters in the subsequent FC layer. This mitigates overfitting. Additionally, GAP provides spatial invariance, making the learned feature representations robust to the locations of input patterns. This is advantageous for handling shifting or impulse noise. Thus, GAP confers the benefits of overfitting reduction and spatial invariance before classification.

\subsection{Modular Semi-Supervised Module}

For an unlabeled data point $x$, prior work has incorporated the consistency regularization term
\begin{equation}
	|| \mathrm{p_{model}}(y|\mathrm{Aug}(x);\theta)- \mathrm{p_{model}}(y|\mathrm{Aug}(x);\theta)||^2_2,
\end{equation}
where $\mathrm{p_{model}}(y|x;\theta)$ denotes the model prediction outputting a distribution over class labels $y$ for the input $x$ with parameters $\theta$, and $\mathrm{Aug}(x)$ represents the augmentation for the input $x$. Since $\mathrm{Aug}(x)$ represents a stochastic augmentation, those two terms of $\mathrm{p_{model}}(y|\mathrm{Aug}(x);\theta)$ are not identical. 
MixMatch \cite{berthelot2019mixmatch} employs consistency regularization by leveraging standard data augmentations for the input signals.

\begin{figure*}
	\centering
	\includegraphics[width=0.95\linewidth]{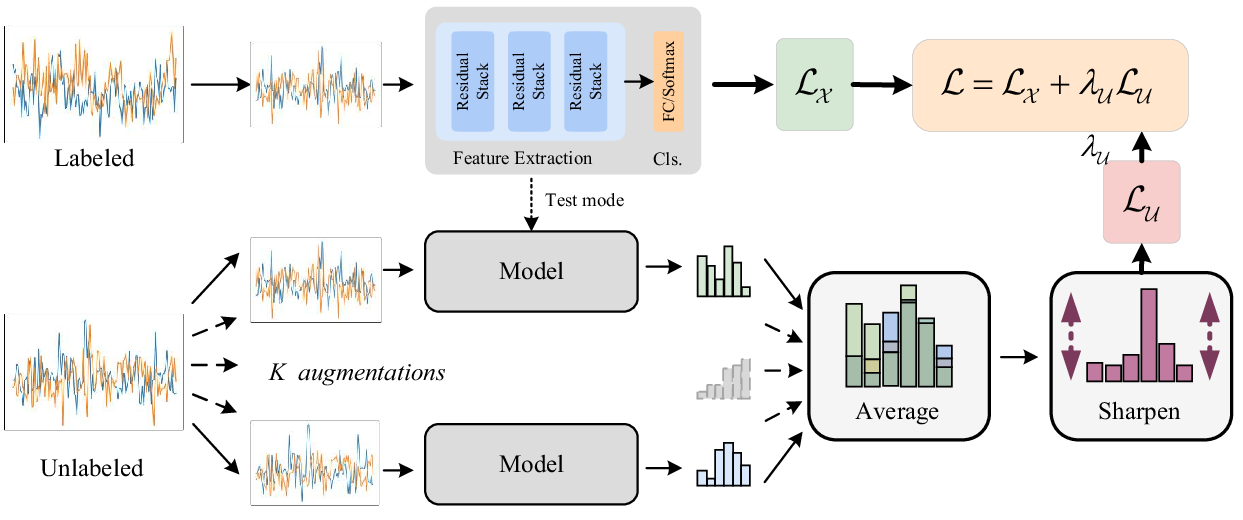} 
	\caption{Illustration of the label guessing process used in MixMatch.}
	\label{fig:mixmatch}
\end{figure*}

MixMatch combines components from multiple prevailing semi-supervised learning paradigms into a holistic approach. Given a batch of labeled examples $\mathcal{X}$ with one-hot encoded labels and an equally-sized batch of unlabeled examples $\mathcal{U}$, MixMatch generates augmented labeled examples $\mathcal{X}^e$ and unlabeled examples $\mathcal{U}^e$ with predicted “guessed” labels. $\mathcal{X}^e$ and $\mathcal{U}^e$ are then used to calculate supervised and unsupervised loss terms, respectively. The overall loss $\mathcal{L}$ for semi-supervised learning in MixMatch is generally defined as
\begin{equation}
	\mathcal{X}^e,\mathcal{U}^e= \mathrm{MixMatch}(\mathcal{X},\mathcal{U},T,K,\alpha),
\end{equation}
where $H(p, q)$ denotes the cross-entropy between the predict distribution $p$ and the target distribution $q$, and $T$, $K$, and $\alpha$ are hyperparameters. Algorithm \ref{alg:mixmatch} presents the full version of MixMatch, and Fig. \ref{fig:mixmatch} illustrates the diagram of the label-guessing process. 

\begin{algorithm} 
\caption{Given a batch of labeled data $\mathcal{X}$ and a batch of unlabeled data $\mathcal{U}$, MixMatch generates an expanded set of labeled examples $\mathcal{X}^e$ and an expanded set of unlabeled examples $\mathcal{U}^e$ with predicted labels.} 
\label{alg:mixmatch} 
\begin{algorithmic}[1] 
\REQUIRE Batch of labeled data and corresponding one-hot labels $\mathcal{X}=((x_b,p_b);b\in (1,\cdots,B))$, batch of unlabeled examples $\mathcal{U}=(u_b;b\in (1,\cdots,B))$, sharpening temperature $T$, number of augmentations $K$, Beta distribution parameter $\alpha$ for MixUp;
\ENSURE $\mathcal{X}^e,\mathcal{U}^e$.
\STATE \textbf{for} $b=1$ \textbf{to} $B$ \textbf{do}
\STATE \quad $\hat{x}_b=\mathrm{Aug}(x_b)$ 
\STATE \quad \textbf{for} $k=1$ \textbf{to} $K$ \textbf{do}
\STATE \quad \quad $\hat{u}_{b,k}=\mathrm{Aug}(u_b)$
\STATE \quad \textbf{end for}
\STATE \quad $\overline{q}_b=\frac{1}{K}\sum_k\mathrm{p_{model}}(y|\hat{u}_{b,k};\theta)$
\STATE \quad $q_b=\mathrm{Sharpen}(\overline{q}_b,T)$
\STATE \textbf{end for}
\STATE $\hat{\mathcal{X}} = ((\hat{x}_b,p_b),b\in (1,\cdots,B))$
\STATE $\hat{\mathcal{U}} = ((\hat{u}_{b,k},p_b),b\in (1,\cdots,B),k\in (1,\cdots,K))$
\STATE $\mathcal{W}=\mathrm{Shuffle}(\mathrm{Concat}(\mathcal{\hat{X},\hat{U}}))$
\STATE $\mathcal{X}^e = (\mathrm{MixUp}(\mathcal{\hat{X}}_i,\mathcal{W}_i);i\in (1,\cdots,|\mathcal{\hat{X}}|))$
\STATE $\mathcal{U}^e = (\mathrm{MixUp}(\mathcal{\hat{U}}_i,\mathcal{W}_{i+|\mathcal{\hat{X}}|});i\in (1,\cdots,|\mathcal{\hat{U}}|))$
\end{algorithmic}
\end{algorithm}

\subsubsection{Data augmentation}
Data augmentation is utilized for both the labeled examples $\mathcal{X}$ and unlabeled examples $\mathcal{U}$. For each labeled example $x_b \in \mathcal{X}$, a transformed version $\hat{x}_b$ is generated by applying a stochastic augmentation function $\hat{x}_b = \mathrm{Aug}(x_b)$ (Algorithm \ref{alg:mixmatch}, line 2). Similarly, each unlabeled example $u_b \in \mathcal{U}$ is augmented $K$ times by drawing $K$ augmented versions $\hat{u}_{b,k} = \mathrm{Aug}(u_b), k \in (1,\dots,K)$ (Algorithm \ref{alg:mixmatch}, line 4). Individual augmentations are used to generate a “guessed label” $q_b$ for each $u_b$, which includes \texttt{RandomFlip} and \texttt{RandomSmooth}. 
\texttt{RandomFlip} refers to the data augmentation method presented in \cite{huang2019data}, which converts the $I$ or $Q$ part to $-I$ or $-Q$ randomly. 
\texttt{RandomSmooth} randomly selects several points and replaces them with the mean of the previous and next values.

\subsubsection{Label guessing}
For each unlabeled data $u$ in $\mathcal{U}$, MixMatch generates a pseudo-label for the data under the prediction mode of the model, which is used in the unlabeled loss term. 
The average of the model's predicted class distributions across all the $K$ augmentations of $u_b$ is computed as 
\begin{equation}
	\overline{q}_b=\frac{1}{K}\sum^K_{k=1}\mathrm{p_{model}}(y|\hat{u}_{b,k};\theta),
\end{equation}
in Algorithm \ref{alg:mixmatch}, line 6. 
In the consistency regularization methods, data augmentation is commonly used to create an artificial target for an unlabeled example.

To reduce the entropy of the label distribution under the average prediction over augmentations $\overline{q}_b$, a sharpening function is applied. Practically, the “temperature” of the categorical distribution is modified, which is classified as the sharpening operation, given as
\begin{equation}
	\mathrm{Sharpen}(p,T)_i:=p_i^{\frac{1}{T}}/ \sum_{j=1}^L p_{j}^{\frac{1}{T}},
\end{equation}
where $p$ is the input categorical distribution ($p$ is the average class distribution over augmentations $\overline{q}_b$ in MixMatch, as in Algorithm \ref{alg:mixmatch}, line 7) and $T$ denotes the hyperparameter. 
When $T \to 0$, the output of $\mathrm{Sharpen}(p,T)$ reaches a Dirac (“one-hot”) distribution. 
$q_b = \mathrm{Sharpen}(\overline{q}_b, T)$ is used as an object for the model’s forecast for augmentation of $u_b$, and thus lower-entropy predictions are generated by the model when the temperature is lower. 

\subsubsection{MixUp}
Both labeled examples and unlabeled examples with predicted labels are utilized to train the model. 
A slightly modified version of MixUp is defined to be compatible with the separate loss terms. 
For a pair of two examples with corresponding label distributions $(x_1, p_1)$, $(x_2, p_2)$, $(\tilde{x}, \tilde{p})$ is computed as
\begin{subequations}
\begin{align}
	\lambda &\sim \mathrm{Beta}(\alpha,\alpha) \\
	\tilde{\lambda} &= \max(\lambda,1-\lambda)\label{eq:12}\\
	\tilde{x} &=\tilde{\lambda} x_1 + (1-\tilde{\lambda})x_2\\
	\tilde{p} &=\tilde{\lambda} p_1 + (1-\tilde{\lambda})p_2
\end{align}
\end{subequations}
where $\alpha$ is a hyperparameter, and Vanilla MixUp sets $\tilde{\lambda}=\lambda$ directly. Labeled and unlabeled training instances are mixed within the same mini-batch, and the order of the batch to calculate individual loss terms should be preserved appropriately. Eq. (\ref{eq:12}) is used to achieve this, which ensures that $\tilde{x}$ is closer to $x_1$ than to $x_2$. To implement MixUp, all augmented labeled training examples along with their ground truth labels are aggregated with unlabeled examples that are assigned estimated labels, given as
\begin{subequations}
\begin{align}
	\mathcal{\hat{X}} &=((\hat{x}_b,p_b);b\in (1,\cdots,B)), \\
	\mathcal{\hat{U}} &=((\hat{u}_{b,k},q_b);b\in (1,\cdots,B),k\in(1,\cdots,K)).
\end{align}
\end{subequations}

Then, these collections are combined by using $\mathrm{Concat}$ operation, and the result is shuffled by using $\mathrm{Shuffle}$ operation to generate $\mathcal{W}$, functioning as the input data for MixUp (Algorithm \ref{alg:mixmatch}, line 11). 
For the $i$th labeled example $\mathcal{\hat{X}}_i$ and its corresponding label $\mathcal{W}_i$ in the augmented labeled set $\mathcal{\hat{X}}$, the MixUp augmentation is applied as $\mathrm{MixUp}(\mathcal{\hat{X}}_i, \mathcal{W}_i)$. The output is then added to the aggregated dataset $\mathcal{X}^e$ (Algorithm \ref{alg:mixmatch}, line 12). 
$\mathcal{U}^e_i=\mathrm{MixUp}(\mathcal{\hat{U}}_i,\mathcal{W}_{i+|\mathcal{\hat{X}}|})$ for $i \in (1,\cdots,|\mathcal{\hat{U}}|)$ is computed by using the remaining part of $\mathcal{W}$ that is not utilized in the construction of $\mathcal{X}^e$ (Algorithm \ref{alg:mixmatch}, line 13). 
MixMatch transfers $\mathcal{X}$ into $\mathcal{X}^e$, a set of labeled examples that are augmented and MixUp. 
Similarly, $\mathcal{U}$ is turned into $\mathcal{U}^e$, which is a collection of multiple-augmented unlabeled samples with corresponding guessed labels. 

\subsubsection{Hyperparameters}
Since MixMatch incorporates multiple strategies for utilizing unlabeled data, it brings several hyperparameters, including the sharpening temperature $T$, number of the unlabeled augmentations $K$, $\alpha$ parameter for Beta in MixUp, and the unsupervised loss weight $\lambda_{\mathcal{U}}$. 
In practice, semi-supervised learning algorithms with multiple hyperparameters can present difficulties, as cross-validation with small validation sets may not provide reliable hyperparameter estimates \cite{rasmus2015semi,oliver2018realistic}. 
However, most of the hyperparameters of MixMatch can be constant and do not need to be tuned during the training process as indicated in \cite{berthelot2019mixmatch}.
Thus, we follow the set for all experiments, where $T=0.5$, $K=2$, $\alpha=0.75$ and $\lambda_{\mathcal{U}}=75$.

\subsection{Loss Function}
The cross-entropy loss is a commonly used objective function for optimization in multi-class classification models \cite{goodfellow2016deep}. Compared to the traditional squared error loss, the cross-entropy loss often results in faster training convergence since the gradients of the model weights are less likely to vanish. To compute the cross-entropy loss, the output of the final layer must first be passed through a softmax function to normalize the predictions to a probability distribution over the classes, as shown in eq. (\ref{eq:ce}). The per-sample cross-entropy loss can then be expressed as
\begin{equation}\label{eq:ce}
	E = -\sum^{C}_{c=1}t_c\log(y_c),
\end{equation}
where $t$ is the target output vector and $t_c$ is the model's predicted probability that a given sample belongs to the class $c$. The parameters of the model can then be optimized to minimize the cross-entropy loss using gradient descent over a number of training iterations.

Given the augmented labeled batch $\mathcal{X}^e$ and unlabeled batch $\mathcal{U}^e$, the semi-supervised loss function defined in eq. (\ref{eq_lx}) to (\ref{eq_l}) is utilized for training. The combination of the cross-entropy loss between labels and model predictions from $\mathcal{X}'$, and the squared L2 loss on predictions and guessed labels from $\mathcal{U}^e$, is expressed in eq. (\ref{eq_l}) along with a bias of $\lambda_{\mathcal{U}}$. The L2 loss is commonly utilized for the unlabeled data to gauge predictive uncertainty due to its bounded nature and reduced sensitivity to incorrect predictions.
Gradient propagation is not performed when computing the guessed labels.

\begin{subequations}
\begin{align}
\mathcal{L}_{\mathcal{X}}&=\frac{1}{|\mathcal{X}^e|} \sum_{x,p\in\mathcal{X}^e}\mathrm{H}(p,\mathrm{p_{model}}(y|x;\theta)),\label{eq_lx}\\
\mathcal{L}_{\mathcal{U}}&=\frac{1}{L|\mathcal{U}^e|} \sum_{u,q\in\mathcal{U}^e}||q-\mathrm{p_{model}}(y|u;\theta)||^2_2,\\
\mathcal{L}&=\mathcal{L_X}+\lambda_{\mathcal{U}}\mathcal{L_U}.\label{eq_l}
\end{align}
\end{subequations}

\section{Experiments}
\label{sec:experiments}

\subsection{Datasets}

Table \ref{tab:datasets} illustrates the datasets used for each experiment and the corresponding results. 
For AMC, we choose the popular data set RadioML.2016.04C \cite{o2016convolutional}, which is provided by DeepSig\footnote{\url{https://www.deepsig.ai/datasets}}. The dataset comprises 8 types of digital signals, namely BPSK, QPSK, 8PSK, QAM16, QAM64, GFSK, CPFSK, and PAM4, along with 3 types of analog signals which are WBFM, AM-SSB, and AM-DSB. By the specifications outlined in \cite{li2020automatic}, we opt for a signal-to-noise ratio (SNR)  range of $-6$ dB to $12$ dB with an interval of $2$ dB in our experiment. The dataset contains a total of $81,030$ samples. The dynamic channel model hierarchical block, which encompasses frequency offset, sample rate offset, additive white Gaussian noise (AWGN), multipath, and fading, is utilized for channel simulation. The signal sequences are aggregated without any designation of a training or a testing set. To ensure that both the training set and the testing set adhere to the same data distribution, we adopt the “hold out” method for selecting distinct samples from the dataset. Specifically, $20\%$ of the dataset is selected for testing by using a stratified sampling method. Corresponding ratios (i.e. $3\%$, $5\%$, or $10\%$) of the dataset are trained with labels, and the remainder is utilized without labels. 

To provide additional evidence of the effectiveness of the proposed approach, we employ a larger variant of the RadioML dataset, namely RadioML.2016.10A \cite{o2016convolutional}, for supervised learning. This version of the dataset is a more refined and standardized version of RadioML.2016.04C and includes $220,000$ labeled data points spanning the same class and SNR range. Specifically, we follow the settings in our previous works \cite{zhang2021novel,zhang2021automatic}. 

To conduct wireless technology classification, we utilize the Technology Recognition (TechRec) dataset \footnote{\url{https://ieee-dataport.org/documents/iq-signals-captured-lte-wifi-and-dvb-t}}  \cite{fontaine2019towards}. This dataset comprises in-phase and quadrature (IQ) signals captured from multiple wireless technologies (LTE, Wi-Fi and DVB-T) deployed in varied environments, and is collected over the air at six distinct locations in Gent, Belgium (UZ, Reep, Rabot, Merelbeke, iGent and Gentbrugge). Our experiments employ a SNR range of $-6$ dB to $12$ dB, with an interval of $2$ dB, and the dataset contains a total of $160,800$ samples.

\begin{table*}[h]
\centering
\caption{Datasets for Wireless Signal Recognition.}
\begin{center}
\begin{tabular}{m{2cm}<{\centering}m{2cm}<{\centering}m{6cm}<{\centering}m{4cm}<{\centering}m{2cm}<{\centering}}
\toprule
Task & Dataset & Classes & Description & Results\\
\midrule
AMC & RadioML 2016.10A \cite{o2016convolutional} & AM-SSB, CPFSK, QPSK, GFSK, PAM4, QAM16, WBFM, 8PSK, QAM64, AM-DSB, BPSK &$220,000$ samples with a SNR range of [-6:2:12]  & Fig. \ref{fig:comparison}, \ref{fig:layers}, \ref{fig:loss}, {\color{blue}Table \ref{tab:parameters}} \\
\midrule
AMC & RadioML 2016.04C \cite{o2016convolutional} & AM-SSB, CPFSK, QPSK, GFSK, PAM4, QAM16, WBFM, 8PSK, QAM64, AM-DSB, BPSK & $81,030$ samples with a SNR range of [-6:2:12] & Fig. \ref{fig:few}, \ref{fig:confusion}, \ref{fig:mixmatchresult}  \\
\midrule
WTC & TechRec \cite{fontaine2019towards} & LTE, Wi-Fi and DVB-T &$160,800$ samples with a SNR range of [-6:2:12]   &Fig. \ref{fig:wtc}, \ref{fig:wtc_few} \\
\bottomrule
\end{tabular}
\label{tab:datasets}
\end{center}
\end{table*}

To train our proposed SSwsrNet framework, we employ the Adam optimization algorithm with an initial learning rate of $0.001$ over $50$ training epochs. When training the model under few-sample conditions, the same initial learning rate is utilized. To assess the performance of various algorithms, model accuracy is employed as the evaluation metric. Accuracy is defined as the ratio of correctly predicted observations to the total observations. Mathematically, it is defined as
\begin{equation}
	acc = \frac{t}{t+f},
\end{equation}
where $t$ is the number of correctly predicted observations and $f$ is the number of wrong observations.

\subsection{Performance Comparison for AMC}

In this subsection, we seek to demonstrate the performance of the proposed SSwsrNet framework under various conditions. First, we use the RadioML 2016.10A dataset \cite{zhang2021novel} to illustrate the performance of our proposed SSwsrNet framework under full supervision. Next, we utilize the RadioML 2016.04C dataset \cite{li2020automatic} to demonstrate the classification accuracy of our proposed model under few-sample conditions. The objectives of these experiments are to illustrate the efficacy of the SSwsrNet approach for both fully-supervised and few-sample learning settings.

\begin{figure}
\centering
\includegraphics[width=0.9\linewidth]{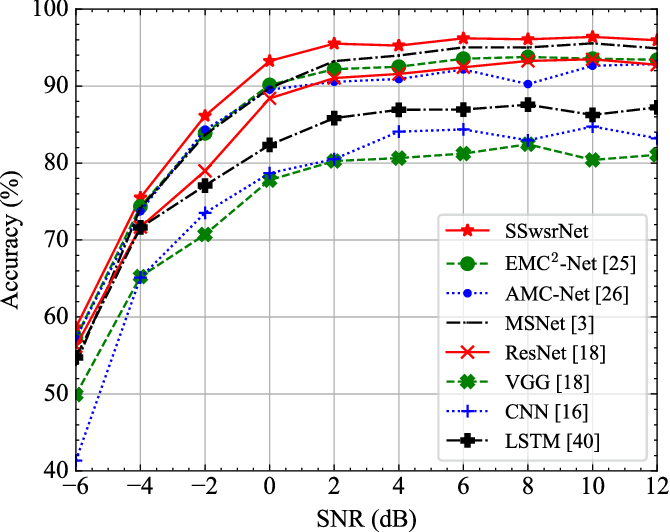} 
\DeclareGraphicsExtensions.
\caption{Classification performance comparison among SSwsrNet, AMC-Net\cite{zhang2023amc}, MSNet \cite{zhang2021novel}, CNN \cite{o2016convolutional}, VGG \cite{o2018over}, ResNet \cite{o2018over} and LSTM \cite{rajendran2018deep} on RadioML 2016.10A \cite{o2016convolutional} with $80\%$ data for training and $20\%$ for testing. }
\label{fig:comparison}
\end{figure}

As shown in Fig. \ref{fig:comparison}, we compare the performance of our proposed SSwsrNet framework with several state-of-the-art DL-based AMC methods including EMC$^2$-Net \cite{ryu2023emc}, AMC-Net \cite{zhang2023amc}, MSNet \cite{zhang2021novel}, CNN \cite{o2016convolutional}, VGG \cite{o2018over}, ResNet \cite{o2018over}, and LSTM \cite{rajendran2018deep}. The network parameters for these models are based on the related works \cite{zhang2021novel,o2016convolutional,o2018over,rajendran2018deep}. To accommodate the input length of 128, only three layers of the relevant modules are used for VGG and ResNet. Our proposed SSwsrNet model achieves the superior performance compared to other advanced DL-based methods, including the recent EMC$^2$-Net \cite{ryu2023emc}, AMC-Net \cite{zhang2023amc} and MSNet \cite{zhang2021novel}. The SSwsrNet achieves over 90\% accuracy when SNR is larger than -1 dB, which is 1 dB more responsive to the EMC$^2$-Net, AMC-Net and MSNet schemes. Moreover, SSwsrNet reaches an accuracy over 95\% when SNR reaches 2 dB, which is approximately 3\% higher than EMC$^2$-Net and AMC-Net, 4\% higher than ResNet and 12\% higher than VGG. These results demonstrate the effectiveness of the SSwsrNet model for AMC.

To further demonstrate the superiority of the proposed SSwsrNet framework, we compare it with the recently advanced DL-based schemes in terms of the number of parameters, training and testing time. As shown in Table \ref{tab:parameters}, the proposed SSwsrNet framework achieves the best performance in terms of the accuracy. The MSNet has the least number of parameters, while it performs worse with higher training and testing time compared to our proposed SSwsrNet due to its multiscale structure. VGG can run faster than other models, but it has the worst performance in terms of the accuracy. Our proposed SSwsrNet achieves a similar testing speed compared to VGG and less training time than recent works such as EMC$^2$-Net \cite{ryu2023emc} and AMC-Net \cite{zhang2023amc}. Overall, the proposed SSwsrNet framework achieves a good balance between the accuracy and the number of parameters as well as the speed.

\begin{table}
\centering
\caption{The number of parameters and speed for different DL-based schemes.}
\begin{center}
\begin{tabular}{m{2cm}<{\centering}m{1.5cm}<{\centering}m{1cm}<{\centering}m{1cm}<{\centering}m{1cm}<{\centering}}
\toprule
Model & Parameters &  Training (s/epoch) & Testing (s/epoch) & Accuracy\\
\midrule
CNN \cite{o2016convolutional} & 13,130,579 & 3.31 &  0.45 &  75.85\% \\
VGG \cite{o2018over} & 160,536 & \textbf{1.46} & \textbf{0.21} & 74.97\% \\
LSTM \cite{rajendran2018deep} & 2,298,379 & \underline{2.46} & 0.31 & 80.65\% \\
ResNet \cite{o2018over}  &  180,619 & 2.57 & 0.48 & 84.98\% \\
EMC$^2$-Net \cite{ryu2023emc} & 282,372 & 3.52 & 0.35 & 86.46\% \\
AMC-Net \cite{zhang2023amc} & 466,144 & 4.29 &  0.35 & 85.41\% \\
MSNet \cite{zhang2021novel} & \textbf{74,955} & 2.95 & 0.32 & \underline{86.90\%} \\
\midrule
SSwsrNet & \underline{90,915} & 2.56 & \underline{0.25} & \textbf{88.88\%} \\
\bottomrule
\end{tabular}
\label{tab:parameters}
\end{center}
\end{table}

\subsection{Performance Comparison for WTC}

\begin{figure}
\centering
\includegraphics[width=0.9\linewidth]{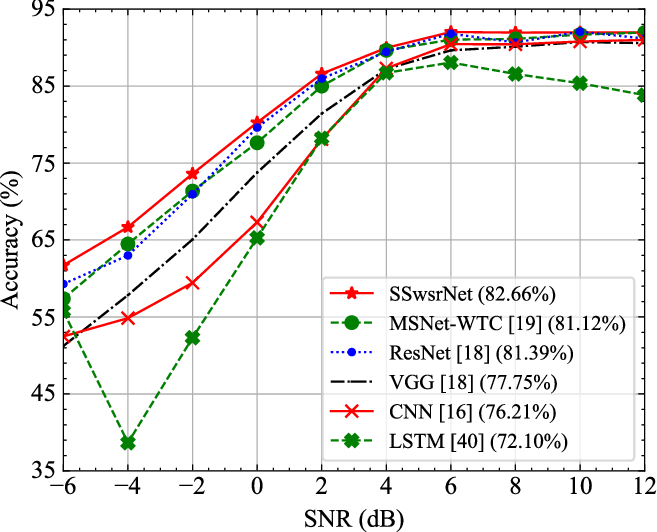} 
\DeclareGraphicsExtensions.
\caption{Classification performance comparison among SSwsrNet,  MSNet-WTC \cite{yuan2021multiscale}, CNN \cite{o2016convolutional}, VGG \cite{o2018over}, ResNet \cite{o2018over} and LSTM \cite{rajendran2018deep} on TechRec \cite{fontaine2019towards} with $80\%$ data for training and $20\%$ for testing. }
\label{fig:wtc}
\end{figure}

In this subsection, to demonstrate the performance of our proposed SSwsrNet for different WSR tasks, simulations are conducted using the dataset TechRec \cite{fontaine2019towards} on the fully supervised situation. As shown in Fig. \ref{fig:wtc}, we compare the proposed SSwsrNet with the recent advanced models including MSNet \cite{yuan2021multiscale}, VGG \cite{o2018over}, ResNet \cite{o2018over}, and LSTM \cite{rajendran2018deep}. We observe that the proposed SSwsrNet achieves the best performance in terms of accuracy compared to other advanced DL-based WTC schemes, including the powerful model MSNet. The SSwsrNet can achieve over $80\%$ $acc$ when SNR is larger than $0$dB, which is $1$ dB more sensitive than the advanced MSNet and ResNet scheme. Moreover, SSwsrNet can reach an accuracy over $90\%$ when SNR reaches $4$ dB, which is $2$ dB more sensitive than VGG and CNN. In terms of the overall classification accuracy, SSwsrNet achieves 82.66\% in all three classes over a wider range of SNRs, which is 1.5\% higher than MSNet and ResNet, and about 5\% higher than VGG and CNN.

\begin{figure}
	\centering
	\includegraphics[width=0.9\linewidth]{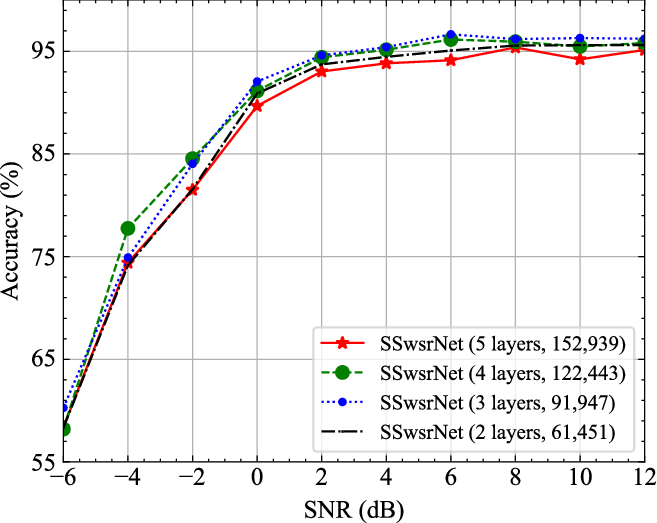} 
	\DeclareGraphicsExtensions.
	\caption{Classification performance over layers on RadioML 2016.10A \cite{o2016convolutional} with $80\%$ training data. }
	\label{fig:layers}
\end{figure}

\begin{figure*}
	\centering
	\subfigure[]{
	\includegraphics[width=0.45\linewidth]{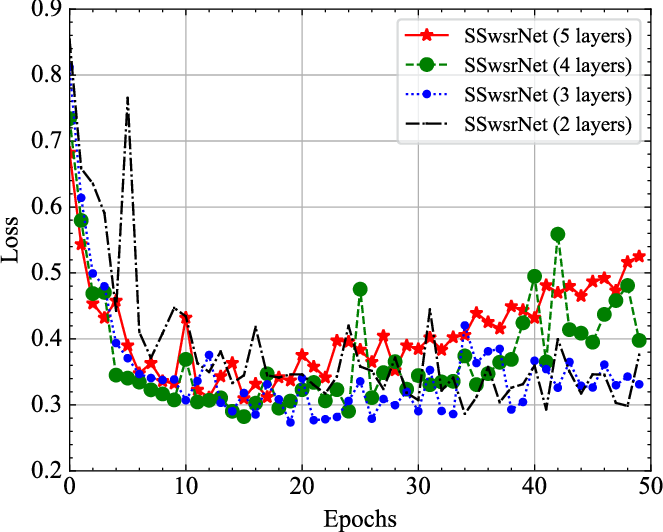} 
	}
	\subfigure[]{
	\includegraphics[width=0.45\linewidth]{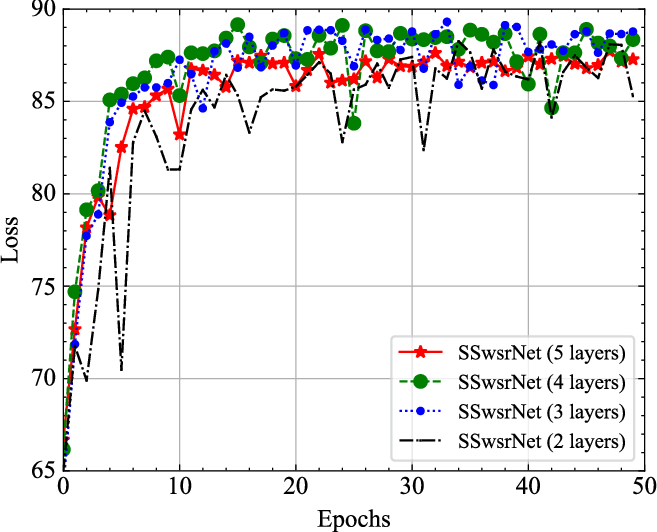} 
	}
	\DeclareGraphicsExtensions.
	\caption{The training process over layers on RadioML 2016.10A\cite{o2016convolutional}. (a) test loss; (b) test accuracy.}
	\label{fig:loss}
\end{figure*}

\begin{figure*}
	\centering
	\subfigure[]{
	\includegraphics[width=0.46\linewidth]{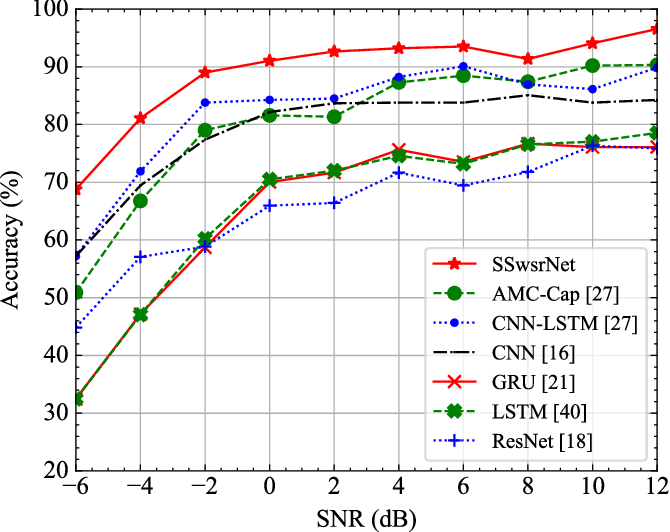} 
	}
	\subfigure[]{
	\includegraphics[width=0.46\linewidth]{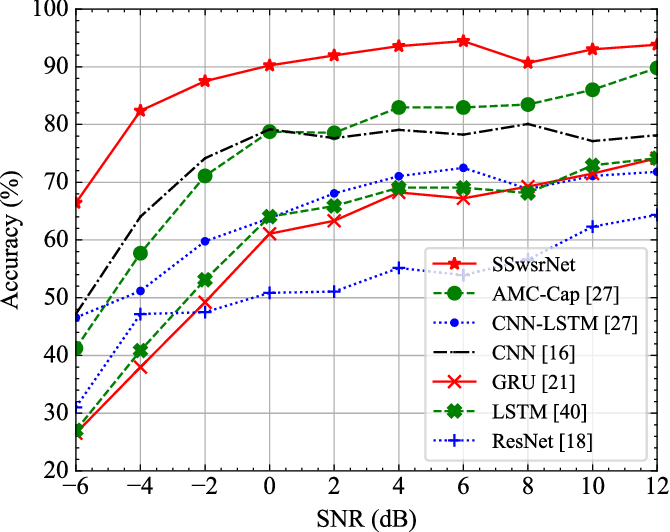} 
	}
	\subfigure[]{
	\includegraphics[width=0.46\linewidth]{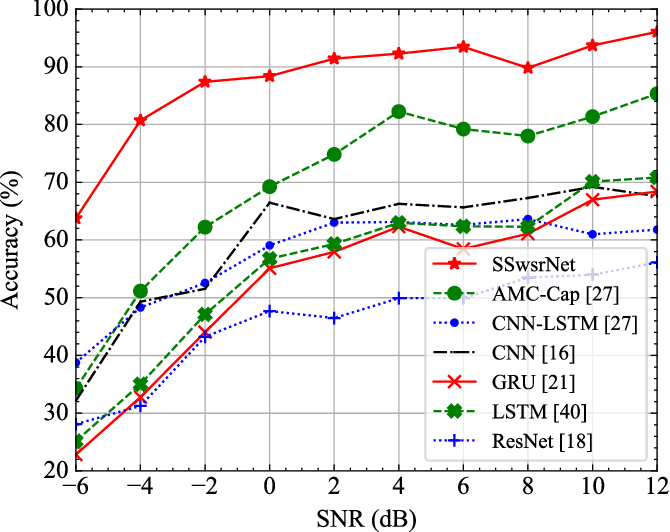} 
	}
	\caption{The impact of the proportion of the labeled training samples on classification accuracy under different models for AMC using RadioML 2016.04C \cite{o2016convolutional}. (a) $10\%$; (b) $5\%$; (c) $3\%$.}
	\label{fig:few}
\end{figure*}

\begin{figure}
	\centering
	\subfigure[]{
	\includegraphics[width=0.9\linewidth]{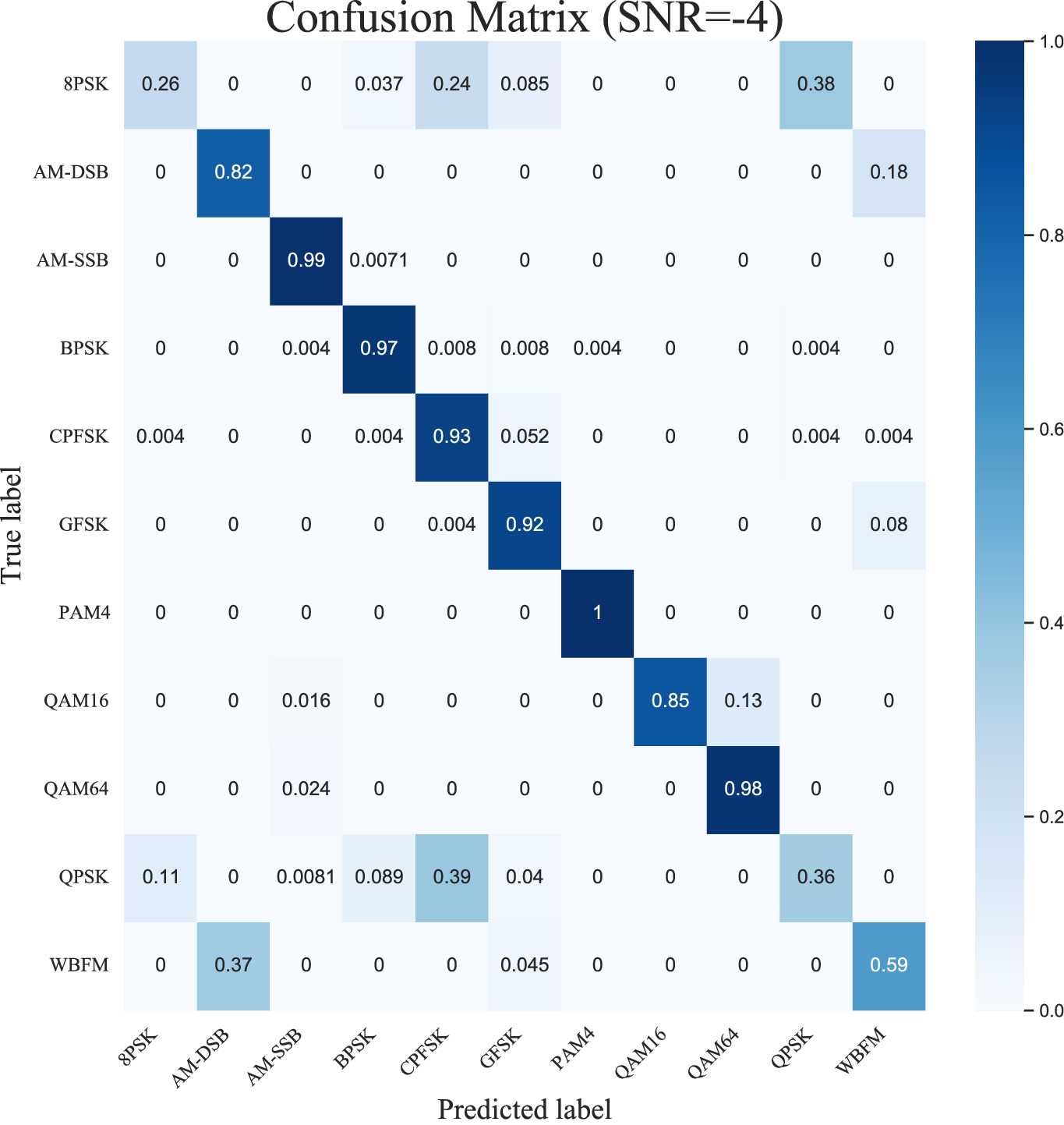} 
	}
	\subfigure[]{
	\includegraphics[width=0.9\linewidth]{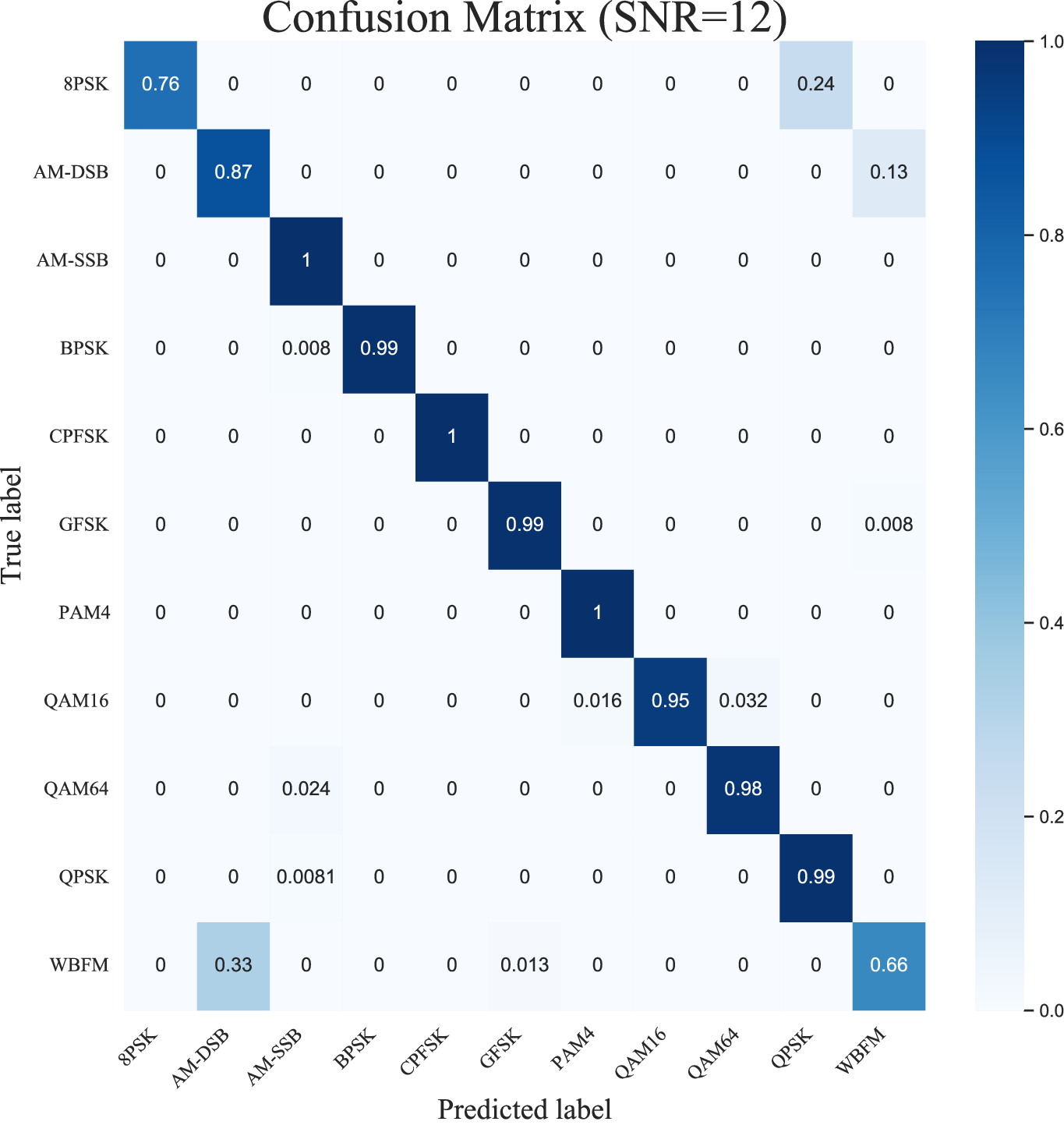} 
	}
	\DeclareGraphicsExtensions.
	\caption{Classification accuracy of modulation signals by SSwsrNet under few samples with $5\%$ labeled data. (a) SNR=$-4$ dB, (b) SNR=$12$ dB.}
	\label{fig:confusion}
\end{figure}

\subsection{Ablation Study}
As demonstrated in the previous works \cite{simonyan2014very,o2018over}, the network depth influences the classification performance of the DL-based models. Thus, we perform an ablation study on the depth of the proposed SSwsrNet. As suggested in \cite{o2018over}, a constant channel number of $32$ is adopted for the SSwsrNet, and the evaluation is applied to the dataset of RadioML 2016.10A \cite{o2016convolutional} with $80\%$ of the data for training. As presented in Fig. \ref{fig:layers}, the performance achieved with different numbers of residual stacks used in the SSwsrNet is given, as well as the corresponding network parameters. It is evident that the SSwsrNet with three layers of residual stacks can achieve the best accuracy among all the number sets. SSwsrNet with two layers of residual stacks has the least network parameters, but its performance is about $2\%$ percent compared to that with three layers in terms of accuracy. To further demonstrate the effectiveness of the SSwsrNet with four layers of residual stacks, we visualize the test loss and test accuracy during the training process. As shown in Fig. \ref{fig:loss} (a) and (b), SSwsrNet with four and five layers of residual stacks causes the over-fitting problem mainly due to their deep architectures with numerous network parameters. Moreover, SSwsrNet with four layers of residual stacks can achieve comparable performance compared to that with three layers. However, four layers bring more trainable parameters and are hard to be trained. Thus, the proposed SSwsrNet is equipped with three layers of residual stacks to reach a balance between accuracy and network parameters.

\subsection{Performance Comparison Under Few-sample Conditions}

In this subsection, we demonstrate the effectiveness of the proposed semi-supervised learning method for WSR under few-sample conditions. To this end, we further investigate the classification accuracy of other network architectures, including CNN\cite{o2016convolutional}, LSTM\cite{rajendran2018deep}, CNN-LSTM \cite{li2020automatic}, GRU\cite{huang2020automatic}, ResNet \cite{o2018over}, and AMR-CapsNet \cite{li2020automatic}, under few-sample settings using the RadioML 2016.04C dataset as in \cite{li2020automatic}. As shown in Fig. \ref{fig:few} (a), when $10\%$ of the dataset is used for training, the classification accuracy of our proposed framework SSwsrNet is close to $95\%$ as the SNR increases. When $5\%$ of the labeled dataset is used for training, the classification accuracy of CNN fluctuates around $80\%$ and AMR-CapsNet can achieve 85\% accuracy when the SNR is greater than 0 dB, while the classification accuracy of SSwsrNet continues to improve by nearly $95\%$. When using $3\%$ of the labeled training data, the classification accuracy of most network structures for modulation signals fluctuates between $60\%$ and $70\%$ and AMR-CapsNet can achieve 80\% accuracy. In contrast, the classification accuracy of SSwsrNet fluctuates around $92\%$. When the SNR is $12$ dB, the classification accuracy is close to $95\%$.

As shown in Fig. \ref{fig:confusion}, $5\%$ of the labeled data ($4,050$ samples) and $75\%$ of the unlabeled data are used to train the SSwsrNet framework, while the remaining data is utilized for model evaluation. We employ a confusion matrix to analyze the performance of modulation signal classification under different SNRs. As it can be seen from Fig. \ref{fig:confusion} (a), at a SNR of $-4$ dB, the model cannot accurately identify the distinguishing characteristics of the modulation signals due to noise, resulting in deviations in determining the signal modulation types. With increasing SNR, the classification accuracy significantly improves for most modulation signals, with most classes achieving over $95\%$ accuracy. However, in line with \cite{li2020automatic}, 8PSK, AM-DSB, and WBFM exhibit lower accuracy. As the SNR increases to $2$ dB in Fig. \ref{fig:confusion} (b), the performance of 8PSK and AM-DSB improves significantly, though WBFM accuracy remains relatively low. Overall, these results demonstrate that the proposed SSwsrNet framework achieves good performance for AMC, especially at higher SNRs.
	
\begin{figure*}
\centering
\subfigure[]{
\includegraphics[width=0.45\linewidth]{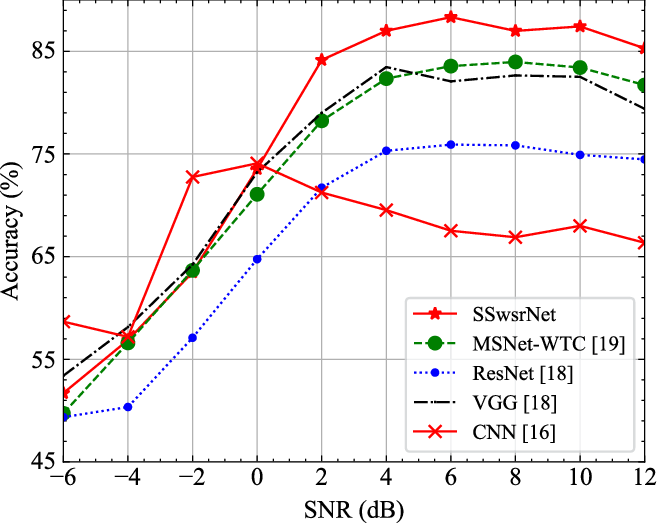} 
}
\subfigure[]{
\includegraphics[width=0.45\linewidth]{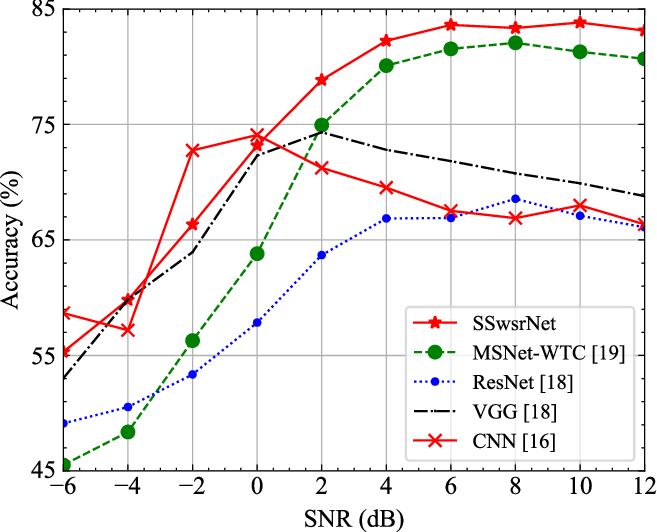} 
}
\subfigure[]{
\includegraphics[width=0.45\linewidth]{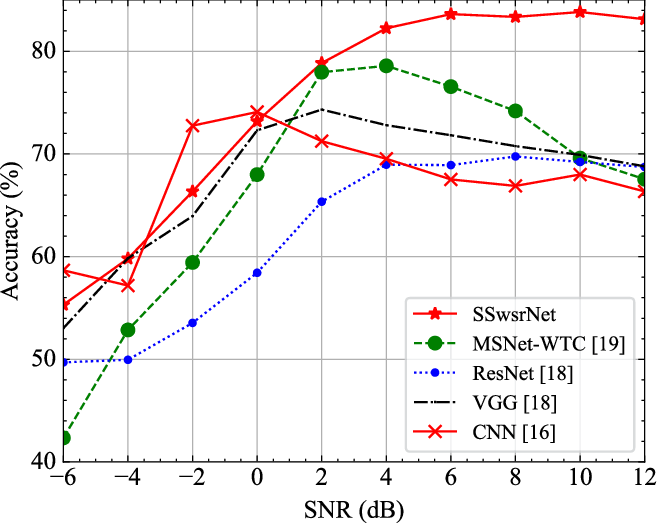} 
}
\subfigure[]{
\includegraphics[width=0.45\linewidth]{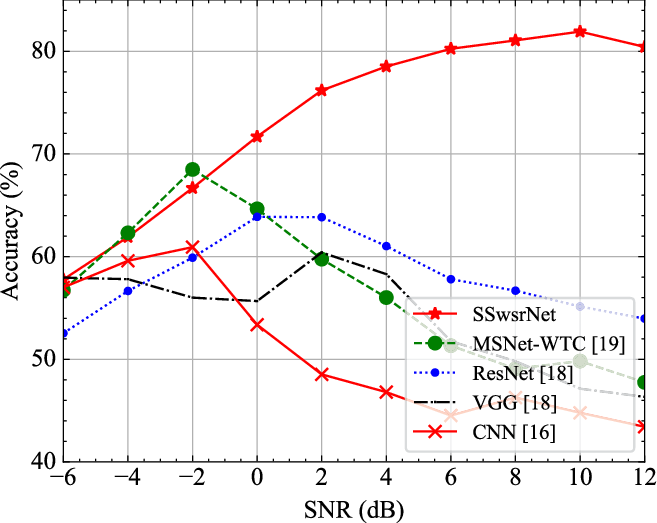} 
}
\DeclareGraphicsExtensions.
\caption{The impact of the proportion of training samples on classification accuracy under different models for WTC using TechRec \cite{fontaine2019towards}. (a) $10\%$; (b) $5\%$; (c) $3\%$; (d) $1\%$. }
\label{fig:wtc_few}
\end{figure*}

Next, we further illustrate the effectiveness of the proposed SSwsrNet and semi-supervised learning method for WTC under limited training data conditions. In Fig. \ref{fig:few}, we compare the proposed SSwsrNet with the recent advanced models including MSNet \cite{yuan2021multiscale}, VGG \cite{o2018over}, ResNet \cite{o2018over}, and LSTM \cite{rajendran2018deep} under different few sample conditions. As shown in Fig. \ref{fig:few} (a), when using $10\%$ of the labeled training dataset, the classification accuracy of our proposed scheme SSwsrNet approaches $90\%$ as the SNR increases, while the classification accuracy of the other advanced DL-based is under 85\%. When $5\%$ of the labeled dataset is used for training, the classification accuracy of MSNet fluctuates around $83\%$ and when the SNR is greater than $4$ dB, while the classification accuracy of SSwsrNet continues to improve by nearly $85\%$. When using $3\%$ of labeled training data, the classification accuracy of most network structures shows similar performance as the labeled data is $5\%$. This is because the amount of labeled data is still quite small. Moreover, we train the models in an extreme few-sample condition with only $1\%$ of labeled data (about $1,286$ samples). As shown in Fig. \ref{fig:wtc_few} (d), training with only $1.3K$ samples, MSNet, ResNet, VGG, and CNN suffer severe overfitting, with worse performance when SNR increases. In contrast, the proposed SSwsrNet with semi-supervised learning can achieve over $80\%$ classification accuracy when SNR is larger than 6dB, which is much better than the compared models.

\subsection{Model Generalization}

\begin{figure}
\centering
\includegraphics[width=0.9\linewidth]{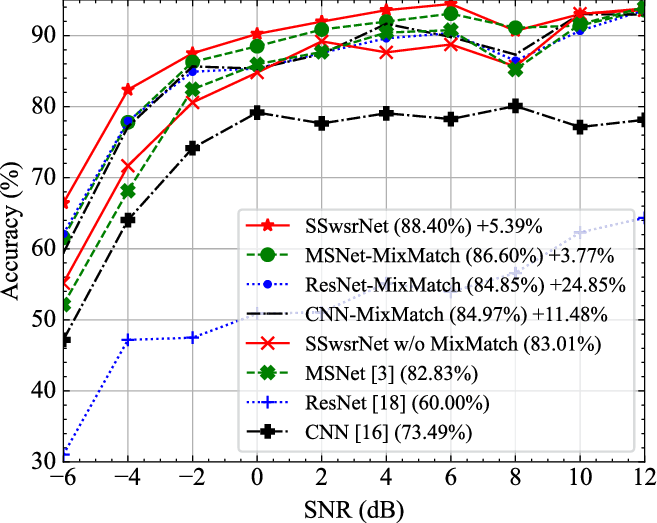} 
\DeclareGraphicsExtensions.
\caption{Classification performance comparison with models with the proposed semi-supervised learning method MixMatch using $5\%$ labeled data trained on RadioML 2016.04C \cite{o2016convolutional}. }
\label{fig:mixmatchresult}
\end{figure}

To demonstrate the generalization of the proposed semi-supervised learning for other DL-based AMC methods under few-sample conditions, we exploit the MixMatch for several DL-based methods including MSNet \cite{zhang2021novel}, ResNet \cite{o2018over}, and CNN \cite{o2016convolutional}. 
As shown in Fig. \ref{fig:mixmatchresult}, we trained all the models by using 5\% labeled data from RadioML 2016.04C. 
The results demonstrate that the proposed SSwsrNet achieves the best performance among all the models under few-sample conditions, with an improvement of +5.39\% over it without MixMatch (\emph{i.e.}, DRSN in isolation). 
SSwsrNet without MixMatch can still achieve the best performance among all the models including MSNet, CNN, and ResNet. 
Moreover, the proposed semi-supervised learning MixMatch can help the conventional DL-based methods such as MSNet, ResNet and CNN perform better under few-sample conditions. 
For example, MSNet with MixMatch can achieve an improvement of $+3.77\%$ over it without MixMatch, achieving about $85\%$ accuracy when SNR is larger than $0$ dB. 
ResNet suffers from severe over-fitting problems under few-sample conditions, and it can achieve comparable performance with the help of MixMatch. 
CNN can achieve about $90\%$ accuracy when SNR is larger than $2$ dB, which is over $10\%$ higher than its original performance.

\begin{figure}
\centering
\includegraphics[width=0.9\linewidth]{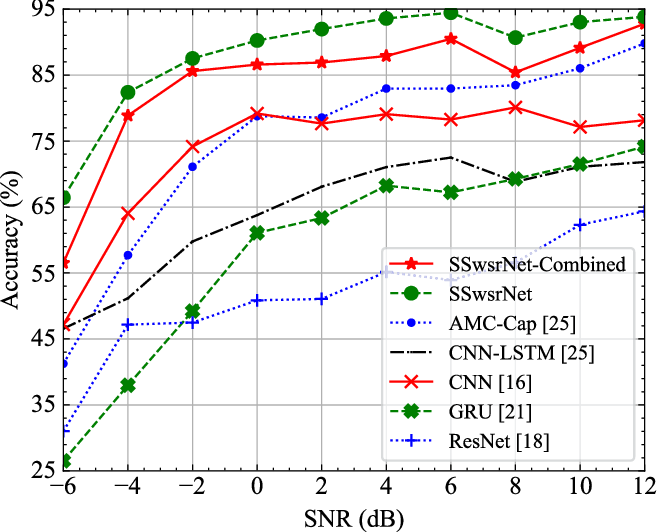} 
\DeclareGraphicsExtensions.
\caption{Classification performance comparison of model generalization using different datasets. }
\label{fig:transfer}
\end{figure}

\begin{figure}
\centering
\subfigure[]{
\includegraphics[width=0.45\linewidth]{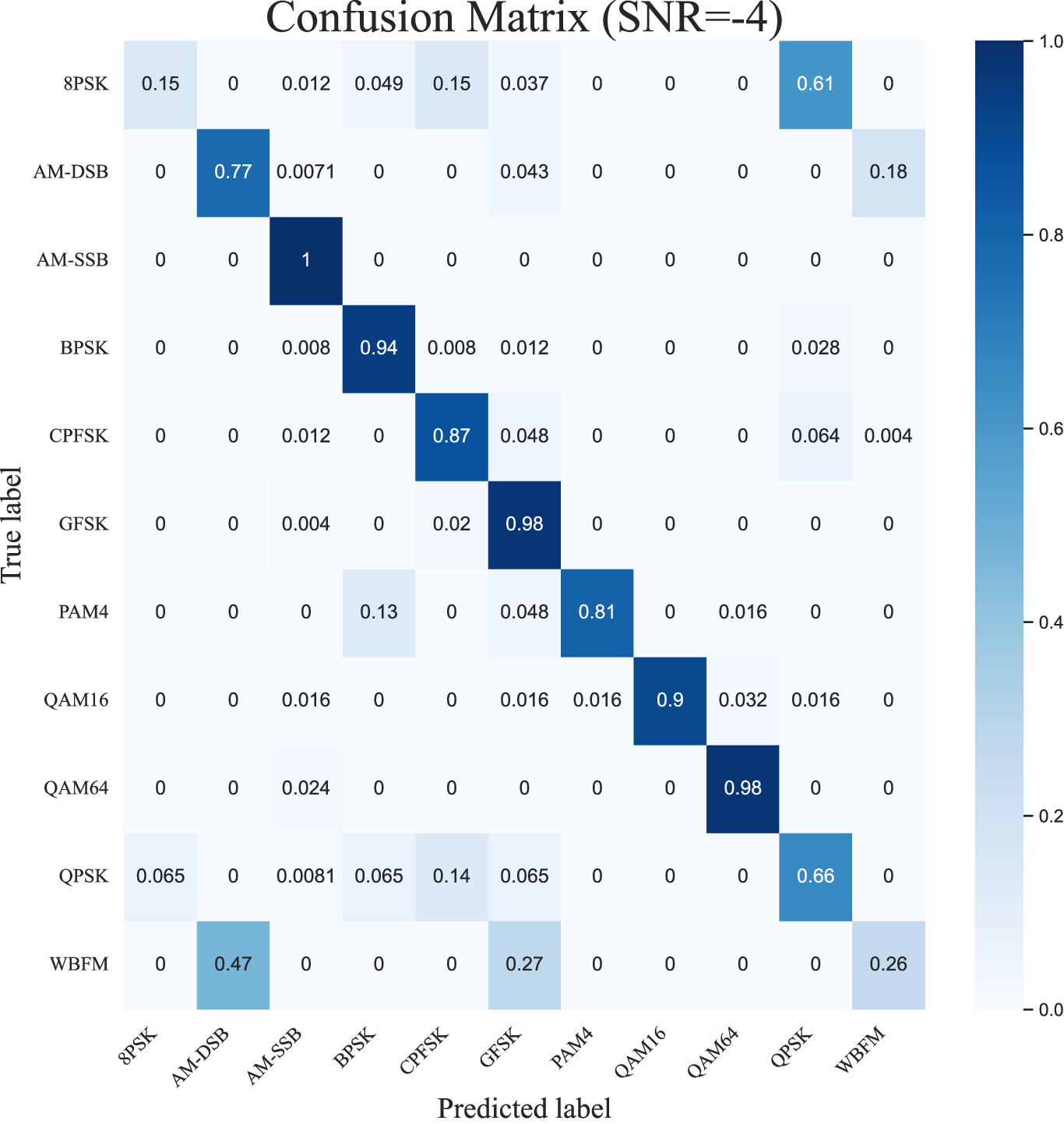} 
}
\subfigure[]{
\includegraphics[width=0.45\linewidth]{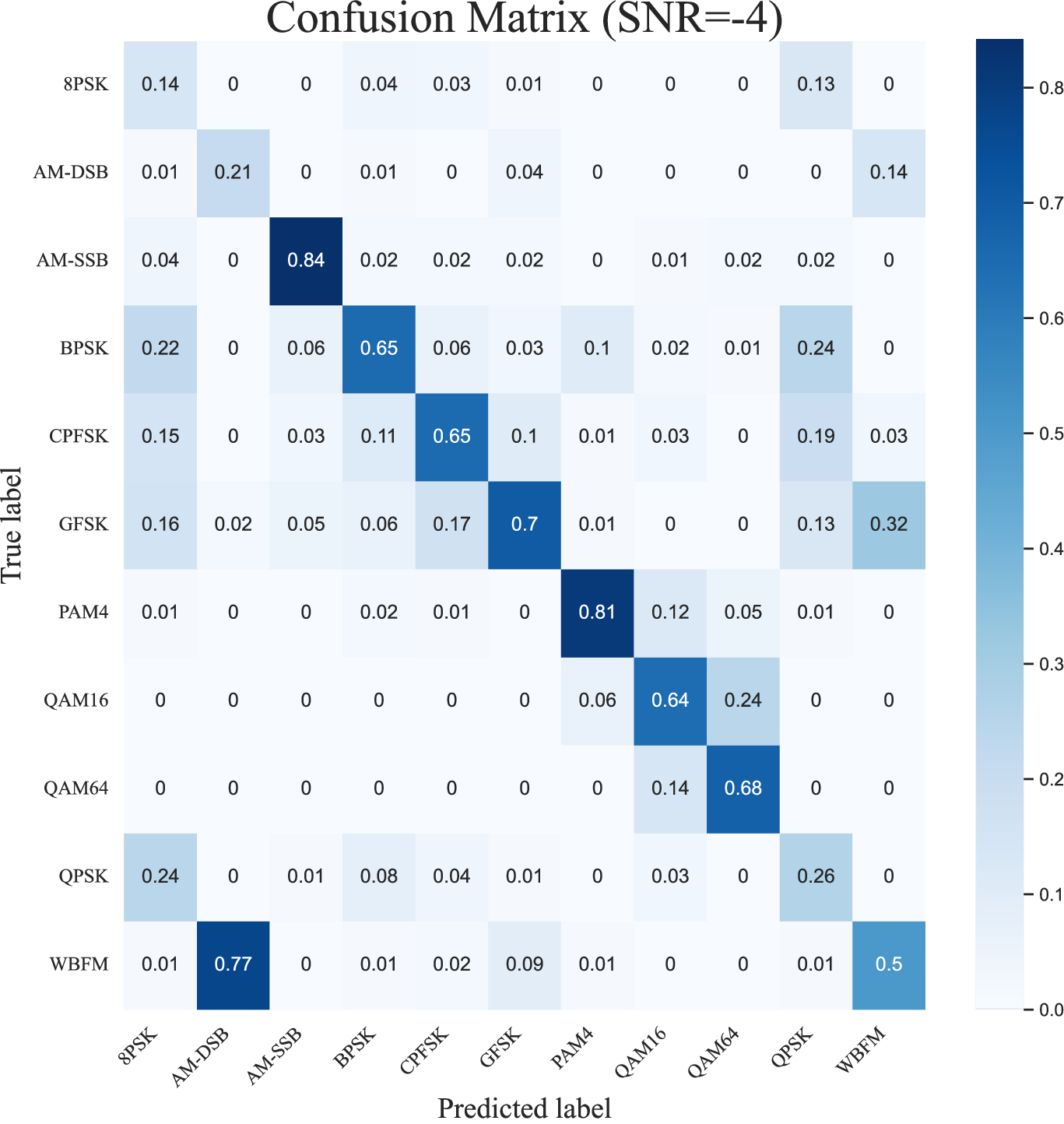} 
}
\DeclareGraphicsExtensions.
\caption{The confusion matrices of (a) SSwsrNet-Combined and (b) AMC-Cap \cite{li2020automatic} under $5\%$ of labeled data of RadioML 2016.04C \cite{o2016convolutional}.}
\label{fig:matrix}
\end{figure}

To further demonstrate the effectiveness of the proposed semi-supervised learning for AMC under few-sample conditions, we evaluate the generalization ability in different datasets. 
Labeled data from RadioML 2016.04C and unlabeled data from RadioML 2016.10A are used for training together. As shown in Table. \ref{tab:datasets}, the labeled data for training and testing is adopted from RadioML 2016.04C with $4,010$ samples and $16,170$ samples, and the unlabeled data is brought from RadioML 2016.10A with $88,000$ samples. The comparison is presented in Fig. \ref{fig:transfer}, where “SSwsrNet-Combined” represents the result of training with different datasets. The combined training can not reach a better performance than that from the same dataset. However, the proposed SSwsrNet under few-sample conditions and combined learning from another dataset can still achieve a promising performance compared to other DL-based methods. Specifically, SSwsrNet-Combined can reach over $85\%$ $acc$ when SNR is larger than $-2$ dB, which is about $4$ dB sensitive than the recent few-shot learning scheme AMC-Cap \cite{li2020automatic}. Moreover, when SNR reaches $12$ dB, SSwsrNet-Combined can obtain a comparable performance compared to SSwsrNet, reaching about $92.7\%$ in terms of accuracy. Fig. \ref{fig:matrix} presents the confusion matrices of SSwsrNet-Combined and AMC-Cap \cite{li2020automatic} under a low SNR of $-4$ dB. Our proposed SSwsrNet can achieve a better performance compared to the AMC-Cap under few-sample conditions with the help of unlabeled data from other datasets. Specifically, $7$ classes in SSwsrNet can achieve an accuracy of $80\%$, while there are only $2$ classes in AMC-Cap. Moreover, the confusion between classes of AMC-Cap is larger than the proposed SSwsrNet.

\section{Conclusion}
\label{sec:conclusion}
In this paper, a novel SSwsrNet framework was proposed by utilizing a deep residual shrinkage network (DRSN) and a modular semi-supervised learning method to address the limited training data issue faced by existing wireless signal recognition (WSR) methods due to the dynamic wireless communication environment. A novel modular semi-supervised learning method is proposed that leverages both labeled and unlabeled data during the model training. Simulation results demonstrated that the proposed framework outperforms benchmark methods in terms of classification accuracy. Additionally, the impacts of network parameters and the effectiveness of the semi-supervised learning approach on the achievable performance were comprehensively investigated. Our proposed SSwsrNet framework can provide meaningful insights for designing DL-based methods addressing WSR with limited labeled data.

\bibliographystyle{IEEEtran}
\bibliography{sswsrnet_ref}

\begin{IEEEbiography}[{\includegraphics[width=1in,height=1.25in,clip,keepaspectratio]{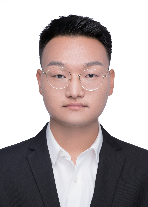}}]{Hao Zhang} (Graduate Student Member, IEEE) received the master's degree from the School of Information Engineering, Nanchang University, China, in 2020. He is currently pursuing the Ph.D. degree in the College of Electronic and Information Engineering, Nanjing University of Aeronautics and Astronautics, Nanjing, China. His research interests focus on machine learning, deep learning, wireless communication, and signal processing.\end{IEEEbiography}

\begin{IEEEbiography}[{\includegraphics[width=1in,height=1.25in,clip,keepaspectratio]{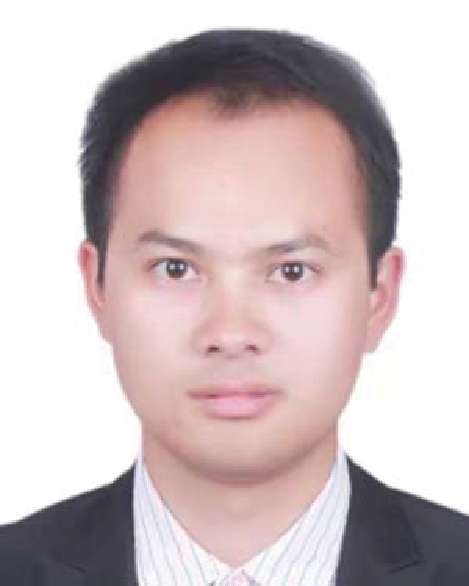}}]{Fuhui Zhou}
(Senior member, IEEE) is currently a Full Professor at Nanjing University of Aeronautics and Astronautics. He is an IEEE Senior Member. His research interests focus on cognitive radio, cognitive intelligence, knowledge graph, edge computing, and resource allocation. He was awarded as IEEE ComSoc Asia-Pacific Outstanding Young Researcher and Young Elite Scientist Award of China and URSI GASS Young Scientist. He serves as an Editor of IEEE Transactions on Communications, IEEE Systems Journal, IEEE Wireless Communications Letters, IEEE Access and Physical Communications.
\end{IEEEbiography}
	
\begin{IEEEbiography}[{\includegraphics[width=1in,height=1.25in,clip,keepaspectratio]{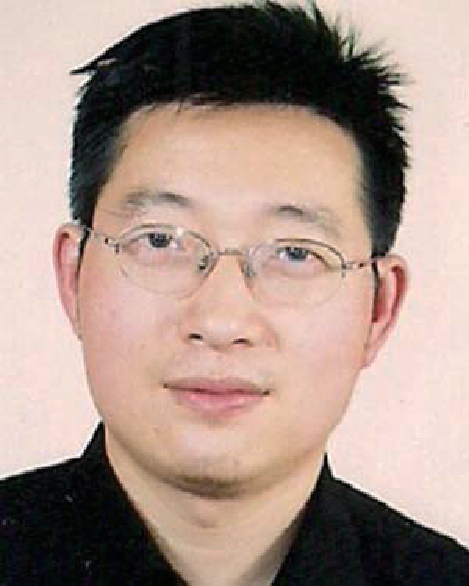}}]{Qihui Wu}
(Fellow, IEEE) received the B.S. degree in communications engineering, the M.S. and Ph.D. degrees in communications and information systems from the Institute of Communications Engineering, Nanjing, China, in 1994, 1997, and 2000, respectively. From 2003 to 2005, he was a Postdoctoral Research Associate with Southeast University, Nanjing, China. From 2005 to 2007, he was an Associate Professor with the College of Communications Engineering, PLA University of Science and Technology, Nanjing, China, where he was a Full Professor from 2008 to 2016. Since May 2016, he has been a Full Professor with the College of Electronic and Information Engineering, Nanjing University of Aeronautics and Astronautics, Nanjing, China. From March 2011 to September 2011, he was an Advanced Visiting Scholar with the Stevens Institute of Technology, Hoboken, USA. His current research interests span the areas of wireless communications and statistical signal processing, with emphasis on system design of software defined radio, cognitive radio, and smart radio.
\end{IEEEbiography}
	
\begin{IEEEbiography}[{\includegraphics[width=1in,height=1.25in,clip,keepaspectratio]{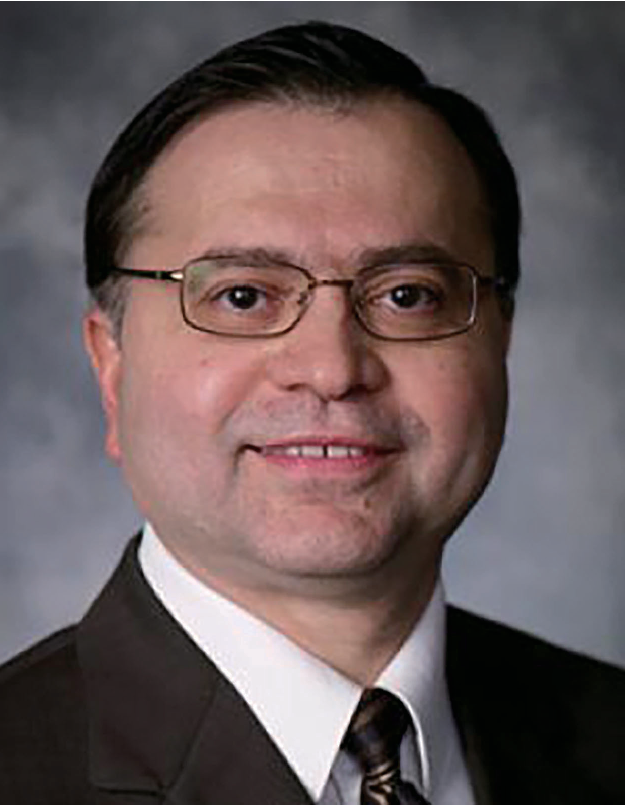}}]{Naofal Al-Dhahir}
(Fellow, IEEE) is Erik Jonsson Distinguished Professor \& ECE Associate Head at UT-Dallas. He earned his PhD degree from Stanford University and was a principal member of technical staff at GE Research Center and AT\&T Shannon Laboratory from 1994 to 2003. He is co-inventor of 43 issued patents, co-author of over 575 papers and co-recipient of 7 IEEE best paper awards. He is an IEEE Fellow, AAIA Fellow, received 2019 IEEE COMSOC SPCC technical recognition award, 2021 Qualcomm faculty award, and 2022 IEEE COMSOC RCC technical recognition award. He served as Editor-in-Chief of IEEE Transactions on Communications from Jan. 2016 to Dec. 2019. He is a Fellow of the US National Academy of Inventors and a Member of the European Academy of Sciences and Arts.
\end{IEEEbiography}

\end{document}